\documentclass[aps, amsfonts, nofootinbib, notitlepage]{revtex4}
\usepackage{epsfig}
\usepackage{amsfonts}
\usepackage{amsmath}
\usepackage{amssymb}

\newcommand{\simge}{\hspace*{0.2em}\raisebox{0.5ex}{$>$}
     \hspace{-0.8em}\raisebox{-0.3em}{$\sim$}\hspace*{0.2em}}
\newcommand{\simle}{\hspace*{0.2em}\raisebox{0.5ex}{$<$}
     \hspace{-0.8em}\raisebox{-0.3em}{$\sim$}\hspace*{0.2em}}

\newcommand{\bea}{\begin{eqnarray}}
\newcommand{\eea}{\end{eqnarray}}

\newcommand{\ket}[1]{| #1 \rangle}                     
\newcommand{\bra}[1]{\langle #1 \, |}                  

\begin{document}

\title{Effective Field Theory for Two-Body Systems with Shallow 
$S$-Wave Resonances} 

\author{J. Balal Habashi}
\affiliation{Department of Physics, University of Arizona, Tucson, AZ 85721, 
USA} 

\author{S. Fleming}
\affiliation{Department of Physics, University of Arizona, Tucson, AZ 85721, 
USA} 

\author{S. Sen}
\affiliation{Department of Physics \& Astronomy, Iowa State University, Ames, 
IA 50011, USA} 

\author{U. van Kolck}
\affiliation{Universit\'e Paris-Saclay, CNRS/IN2P3, IJCLab, 91405 Orsay, France}
\affiliation{Department of Physics, University of Arizona, Tucson, AZ 85721, 
USA} 

\begin{abstract}
Resonances are of particular importance to the scattering of composite 
particles in quantum mechanics. We build an effective field theory 
for two-body scattering which includes a low-energy $S$-wave resonance. 
Our starting point is the most general Lagrangian with short-range interactions.
We demonstrate that these interactions can be organized into various orders
so as to generate a systematic expansion for an $S$ matrix with two
low-energy poles. The pole positions are restricted by renormalization 
at leading order, where the common feature is a non-positive effective range.
We carry out the expansion explicitly to next-to-leading order
and illustrate how it systematically accounts for the results of a toy 
model --- a spherical well with a delta shell at its border.
\end{abstract}

\maketitle

\section{Introduction}

In quantum mechanics, observables are encoded in the $S$ matrix or,
equivalently, in the $T$ matrix.
In the complex momentum plane, poles of the $S$ matrix lying on the positive 
imaginary axis with positive residue correspond to bound states 
\cite{Moller:1946},
while poles on the negative imaginary axis are 
negative-energy virtual states, which are not normalizable.
Pairs of poles can also appear at complex momentum with equal,
negative imaginary parts and equal but opposite real parts. 
These poles can be thought of as resonance states with a finite lifetime
\footnote{We refer to a pole at complex (neither real nor purely imaginary)
momentum as a ``resonance'', regardless of whether it generates a bump
in a cross section.},
which affect scattering significantly when they lie close to the real axis.
Resonances are common in the nonrelativistic scattering 
of atoms and nuclei, sometimes appearing in the
$S$ wave (see, for example, Ref. \cite{TaylorScattering}).
Shallow resonances can lead to significant variation
of phase shifts at low energies.
Examples include a very low-energy resonance in proton-proton scattering 
\cite{Kok:1980dh}
and the $^8\text{Be}$ ground state in the scattering of two alpha particles
\cite{AFZAL:1969zz}.
Here we develop a systematic treatment of shallow 
$S$-wave resonances, or more generally two shallow poles, based
on effective field theory (EFT). 

EFTs exploit a separation of scales
to generate a controlled expansion of observables in 
the small ratio(s) of these scales.
The EFT framework was originally formulated in particle
physics to allow a perturbative treatment of low-energy processes
involving strong-interacting particles \cite{Weinberg:1978kz},
and immediately used to justify the successes of the 
perturbative Standard Model despite the wide range of possibilities
for its underlying dynamics \cite{Weinberg:1979pi}.
In the 1990s, as it was being applied to nuclear physics
(for a review, see for example Refs. \cite{Bedaque:2002mn,Hammer:2019poc}), 
the need arose for EFT to be extended to
shallow, nonrelativistic bound and virtual states, 
which cannot be treated in perturbation theory.
Originally motivated by the two-nucleon problem,
which exhibits a bound state in one $S$ wave and
a virtual state in another,
an EFT description of shallow $S$-wave two-body states was achieved in the 
late 1990s \cite{vanKolck:1997ut,Kaplan:1998tg,Kaplan:1998we,vanKolck:1998bw}.
This was possible using either momentum-dependent contact interactions
or a ``dimeron'' auxiliary field \cite{Kaplan:1996nv}, 
which gives rise to energy-dependent interactions.
Within a few years, shallow two-body resonances also yielded to an EFT
approach, but only with a ``dimeron'' field
\cite{Bertulani:2002sz,Bedaque:2003wa,Higa:2008dn,Gelman:2009be,
Alhakami:2017ntb}.
While the corresponding energy-dependent interactions could be used in 
three-body calculations \cite{Rotureau:2012yu,Ji:2014wta,Ryberg:2017tpv}, 
they are not easily incorporated in most {\it ab initio} methods that enable 
the solution of the Schr\"odinger (or equivalent) equation 
for more than three particles. 
Our goal here is to extend the EFT of 
shallow $S$-wave two-body resonances to
momentum-dependent interactions, which should find wider use.

To understand two-body resonant scattering at low energy in an EFT framework 
we need to identify the characteristic momentum scales inherent to the 
process~\footnote{We use units such that $\hbar = c =1$, so that
mass, momentum, energy, inverse distance, and inverse time all have
the same dimensions.}. 
We consider the scattering of two particles at a momentum $k \sim M_{lo}$
which is much smaller than the inverse range of the interaction $1/R\sim M_{hi}$.
In the nonrelativistic regime, this problem is equivalent
to the the scattering of a particle by a potential of finite range,
but one in which the particle cannot resolve the details of the potential.
It is sensible to make a ``multipole'' expansion of the potential,
which can be considered as a sum of 
Dirac delta functions with an increasing number of derivatives.
Such a potential can be formulated in terms of the most general
Lagrangian density involving only the fields associated with
the particles under study.
The two-body potential is generated
by interactions that involve four fields at the same spacetime point
and their derivatives.
The coefficient of an operator with $n$ derivatives 
--- called Wilson coefficient
in the particle physics literature and low-energy constant (LEC) in
hadronic and nuclear physics --- is a real number $4\pi C_n/m$ 
that encodes the details of the short-range dynamics.
Here we consider for simplicity 
a single particle species of mass $m$ 
and the most common case where
the underlying dynamics is not only Lorentz invariant,
but also symmetric under spatial parity and time reversal.
We also neglect spin degrees of freedom, which bring
no essential complications. In this case only
interactions with an even number of derivatives appear.
The more general case can be obtained straightforwardly
following the same steps as we do below.

We show in the following how scattering around a resonance can be 
described systematically in a series in powers of $M_{lo}/M_{hi}$,
regardless of the details of the underlying interaction,
as long as its range is short compared to the magnitude of the inverse momentum
of the resonance.
The two-body dynamics at low energies is obtained
via the selective resummation of Feynman diagrams, or equivalently by solving
the Schr\"odinger equation.
In either case, as is usual in field theory,
one has to impose a regularization procedure
to cope with the singularity of the interactions.
The conceptually simplest regularization, which we employ below,
imposes a momentum cutoff $\Lambda$.
The choice of regulator is arbitrary and thus makes for a particular
model of the short-range dynamics. 
To ensure observables are insensitive to this arbitrary choice
and model independent, the theory has to be renormalized. 
The $\Lambda$ dependence of each ``bare'' LEC
that appears in the Lagrangian, $4\pi C_n(\Lambda)/m$, is determined by imposing
that a low-energy observable be $\Lambda$ independent.
Other low-energy observables then depend on $\Lambda$
only through positive powers of $1/\Lambda$ and approach
finite limits as $\Lambda \gg M_{hi}$.
The expansion is organized in such a way that renormalization 
holds at each order,
up to terms that have the same magnitude as higher-order
terms when $\Lambda \simge M_{hi}$.
Thus, we can reproduce the effects of any potential exhibiting a 
shallow resonance to arbitrary accuracy. 

The choice of low-energy observables used in the renormalization procedure
is equally arbitrary, up to higher-order terms.
A particularly simple choice is offered by the 
effective-range expansion (ERE) \cite{Bethe:1949yr}.
From general considerations it is possible to show \cite{vanKolck:1998bw}
that an EFT for short-range interactions leads to 
a $T$ matrix for $S$-wave scattering at on-shell momentum 
$k\ll R^{-1}$ which can be expressed in the ERE form 
\begin{equation}
T_0(k) = -\frac{4\pi}{m}\left(k\cot\delta_0(k)-ik\right)^{-1} 
= -\frac{4\pi}{m}\left[-\frac{1}{a_{0}} + \frac{r_{0}}{2}\,k^2 
  - P_{0} \left(\frac{r_{0}}{2}\right)^3 k^4 + \ldots - i k \right]^{-1},
\label{eq.1}
\end{equation}
where $\delta_0(k)$ is the phase shift, and
$a_0$, $r_0$, $P_0$, $\dots$ are
known as, respectively, scattering length, effective range, 
shape parameter\,\footnote{In much of the literature the coefficient of $k^4$ 
is defined to be $P_0 r_0^3$. With our choice $P_0$ takes on more natural values.
},
{\it etc.} 
The values of the ERE parameters can be extracted from the phase
shifts, which in turn can be obtained from
data with only mild theoretical input. 
As is done in much of the EFT literature \cite{Bedaque:2002mn,Hammer:2019poc},
we fix the LECs by forcing them to agree with the 
empirical values of these ERE parameters: $C_0$ is related to
the scattering length, $C_2$ to the effective range, {\it etc.}

A crucial aspect of EFT is ``power counting'' --- the argument
that justifies the expansion of the amplitude in powers of $M_{lo}/M_{hi}$
or ``orders'':
leading order (LO), next-to-leading order (NLO), 
next-to-next-to-leading order (N$^2$LO), and so on.
The simplest assumption is that of ``naturalness'', where 
the size of observables is set solely by the 
large momentum scale $M_{hi}$:
$|a_0| \sim 1/M_{hi}$, 
$|r_0| \sim 1/M_{hi}$, 
$|P_0|\sim 1$, {\it etc.}
(For a review, see Ref. \cite{vanKolck:2020plz}.)
This is assured if the renormalized LECs scale as $C_n={\cal O}(1/M_{hi}^{n+1})$.
In this case $T_0(k)$ is purely perturbative,
with LO consisting of $C_0$ in first order in perturbation theory,
NLO of $C_0$ in second order,
N$^2$LO of $C_0$ in third order and $C_2$ in first order, and so on.
(Starting at N$^2$LO contributions to higher waves are also present.) 
Poles characterized by momentum $|k|\ll M_{hi}$ thus require a certain
amount of fine tuning in the underlying theory such that at least
one interaction is large enough to demand a nonperturbative treatment.

It is relatively easy to find examples of a fine tuning
that produces a large scattering length, $|a_0|\sim 1/M_{lo}\gg 1/M_{hi}$.
This situation corresponds to a single, shallow $S$-wave bound or virtual
state near threshold. This is an intrinsically quantum-mechanical
phenomenon: the particles in the bound state are on average
at distances much larger than the range of the interaction,
which in classical physics determines the size of orbits.
For example, for a square well of fixed range $R$, one can
produce such states at values of the depth $\beta^{2}/mR^{2}$ 
for which $\beta$ is close to an odd multiple of $\pi/2$.
In this case, the LECs scale as 
$C_0={\cal O}(1/M_{lo})$ and $C_{n\ge 2}={\cal O}(1/M_{lo}^2M_{hi}^{n-1})$
\cite{vanKolck:1997ut,Kaplan:1998tg,Kaplan:1998we,vanKolck:1998bw}.
For $k\sim 1/|a_0|$, the interaction with no derivatives and
coefficient $C_0$ is as
important as the unitarity term $ik$, and it needs to be treated
nonperturbatively. All remaining terms in the ERE can
still be treated as perturbations in a distorted-wave Born expansion.
The resulting $S$-wave $T$ matrix is an expansion of Eq. \eqref{eq.1}.
It yields either a bound state or a virtual state, depending 
on the sign of the scattering length, at 
$k \simeq i/a_0 \sim \pm iM_{lo}$. 
This is exactly what happens in nucleon-nucleon scattering at low energies. 
Just as in the natural case, higher waves appear at higher orders.

Neither of these two power-counting schemes can produce a shallow resonance. 
For this to occur we need, in general, to have two fine tunings,
with $|a_0| \sim 1/M_{lo}$ and $|r_{0}| \sim 1/M_{lo}$.
We expect the physics of such a scaling to be produced by 
an effective Lagrangian where 
the two leading operators in the derivative 
expansion are of a size much larger than what is expected from 
naturalness and hence have to be treated nonperturbatively. 
In contrast to other cases, 
$C_0={\cal O}(1/M_{lo})$ and $C_2={\cal O}(1/M_{lo}^3)$, with
other LECs smaller, being suppressed by 
powers of $M_{hi}$. We will argue that
$C_{n\ge 4}={\cal O}(1/M_{lo}^{n/2+2}M_{hi}^{n/2-1})$.
These are the central ideas of our paper. 

Proper renormalization is the cornerstone of our approach.
This is not the first time that the two leading contact 
interactions are solved exactly.
In fact our LO solution reproduces the amplitude of 
Beane, Cohen, and Phillips \cite{Phillips:1997xu,Beane:1997pk}.
These authors have shown that renormalization requires 
$r_0\simle 1/\Lambda$
for $\Lambda\gg  M_{lo}$, which is 
an example of Wigner's bound on phase shifts \cite{Wigner:1955zz}. 
However, in those early days of nuclear EFT, Beane {\it et al.} 
were concerned with bound states.
The fact that two-nucleon data require $r_0> 0$ led those authors to conclude
that EFT was of little use in a nonperturbative context. 
Later, it was shown  
\cite{vanKolck:1997ut,Kaplan:1998tg,Kaplan:1998we,vanKolck:1998bw} 
that in the case of a single shallow
bound or virtual state the two-derivative interaction should not
be iterated to all orders 
\footnote{For a recent proposal on how to iterate corrections
in limited cutoff ranges 
while retaining the EFT expansion, see Ref. \cite{Beck:2019abp}.}
and, when treated perturbatively,
it can accommodate either sign of $r_0$.
Like Beane {\it et al.},
we take renormalizability and its
constraint $r_0\le 0$ seriously.
However, here we note
that we are dealing with a valid EFT ---
just not one with a single fine tuning as relevant in the two-nucleon problem,
but the case where the underlying theory
has the two fine tunings needed for an $S$-wave resonance
or, more generally, two low-energy poles.

With this interpretation, we use the power counting outlined above 
to predict the positions of poles  
with controlled errors, which can be improved by adding 
higher-order operators perturbatively to the LO effective Lagrangian. 
The renormalization requirement $r_0\le 0$ forces resonance
poles to be in the lower half of the complex-momentum plane,
which guarantees their role as decaying states.
More generally, there are two $T$-matrix poles, 
the positions of which are determined in the complex momentum plane 
by the relative sizes of $a_0$ and $r_0\le 0$. The two poles can 
lie 
\begin{itemize}
\item 
both below the real axis with non-vanishing, equal and opposite real parts 
--- a resonance; 
\item 
on top of each other on the negative imaginary axis 
--- a double virtual pole;
\item 
both on the negative imaginary axis, but separated 
--- two virtual states; 
\item 
one on the negative imaginary axis and the other at the origin
--- a virtual state and a ``zero-energy resonance'';
or
\item one on the negative and the other on the positive imaginary axis 
--- one virtual and one bound state. 
\end{itemize}

At NLO, the four-derivative
contact interaction enters, which leads to the shape parameter
$|P_0|\sim M_{lo}/M_{hi}$.
We demonstrate the systematic character of our approach
by considering an example where the underlying interaction
takes a particularly simple form: a spherical well of
range $R$ with a delta-shell potential at its edge. 
By varying the strengths of the two
components of the underlying potential at fixed $R$ 
we can produce two poles in each of the arrangements mentioned above.
Fixing the EFT parameters from the ERE parameters that appear
at each order, we show how the toy-model phase shifts and
pole positions are approximated with increasing accuracy 
when we go from LO to NLO.
In the future we hope to include the Coulomb interaction as well,
so as to be able to consider not only a toy model,
but also an $S$-wave resonance of phenomenological
interest such as the $^8\text{Be}$ ground state.

The organization of the paper is as follows. In Sec. \ref{TBS} we 
show the form of the relevant effective Lagrangian in the derivative expansion,
as well as the potential in LO and NLO. 
In Sec. \ref{seE} we derive the scattering amplitude for the EFT at LO
using the Schr\"odinger equation,
followed by the nonperturbative renormalization of the bare parameters 
$C_0(\Lambda)$ and $C_2(\Lambda)$ of the two leading operators in the 
effective Lagrangian. 
We then find the scattering amplitude for the EFT at NLO 
and perturbatively renormalize the amplitude
using $C_4(\Lambda)$, the bare parameter
of the four-derivative contact interaction. 
(An alternative derivation of the results of this section
using field theory is offered in App. \ref{AppxA},
while some details of the renormalization procedure are given
in  App. \ref{AppxB}.)
In the following section, Sec. \ref{poles},
we demonstrate that our EFT successfully reproduces shallow 
resonant states in the ERE. 
We also analyze the sensitivity 
of the positions of the poles for bound and virtual states to perturbations, 
and the error in the position of the poles is estimated.
We compare in Sec. \ref{toy} the results from EFT with those of a toy model 
which among other states includes two shallow poles.
We conclude in Sec. \ref{conc}.

\section{EFT for Two-Body Scattering}
\label{TBS}

In this section we construct an EFT for scattering that exhibits a shallow 
$S$-wave resonance. 
The energy of scattering particles is restricted to be on the order 
of the resonance energy, assumed to correspond to momentum
much smaller than the inverse of the underlying interaction range.
The scattering process is assumed to be elastic,
meaning there is a single open channel with
particles taken to be in their ground states both before and after 
scattering.
(For scattering with more open channels, see Ref.~\cite{Cohen:2004kf}.) 
For simplicity we limit ourselves to one particle species not affected
by the exclusion principle,
but generalization is straightforward.
As a consequence, we are concerned with a single ``heavy'' 
field \cite{Georgi:1990um} $\psi$,
which encodes the annihilation of a particle,
and contact operators in the Lagrangian that 
are Hermitian, as there is no source or sink.
Furthermore, our EFT is invariant under Lorentz 
(in the form of reparameterization invariance \cite{Luke:1992cs}), 
parity and time-reversal transformations, and conserves 
particle number. 
The most general Lagrangian for scattering without spin 
which respects these symmetries and constraints 
can be written in the form  \cite{Fleming:1999ee}
\bea
\mathcal{L} &=& 
\psi^{\dagger} \left(i\frac{\partial}{\partial t}
+\frac{\overrightarrow{\nabla}^2}{2 m}+ \ldots \right)\psi -\frac{4\pi}{m}
\biggl\{C_{0} \left( \psi \psi \right)^{\dagger} \left( \psi \psi \right) 
- \frac{C_{2}}{8}\left[\left( \psi \psi \right)^{\dagger} 
\left(\psi \overleftrightarrow{\nabla}^2 \psi \right) + \mathrm{H.c.} \right]
\nonumber\\
&&
+\frac{C_{4}}{64}
\left[\left( \psi \psi \right)^{\dagger}
\left(\psi \overleftrightarrow{\nabla}^4\psi \right) + \mathrm{H.c.}
+ 2\left( \psi\overleftrightarrow{\nabla}^2\psi \right)^{\dagger}
\left( \psi\overleftrightarrow{\nabla}^2\psi \right) \right]
+ \ldots 
\biggr\}, 
\label{eq.2}
\eea
where 
$\overleftrightarrow{\nabla} \equiv 
\overrightarrow{\nabla}-\overleftarrow{\nabla}$. 
Here we display explicitly only terms contributing to
the (on-shell) $S$-wave scattering of two particles:
the term quadratic in the field $\psi$ is the nonrelativistic kinetic term, 
while the others correspond to interactions with LECs $4\pi C_{n}/m$. 
The ``$\ldots$'' represent 
operators with more derivatives and fields,
and/or operators that only contribute to two-body scattering off shell
and in higher partial waves.
For example, there is an independent operator with four derivatives,
which vanishes when the two-body system is on shell \cite{Stetcu:2010xq},
which allows us to take the coefficient of 
$(\psi\overleftrightarrow{\nabla}^2\psi)^{\dagger}
(\psi\overleftrightarrow{\nabla}^2\psi)$ 
to be $-\pi C_{4}/8 m$ \cite{Fleming:1999ee}.
Similar choices can be made for higher-derivative operators.
Interactions that vanish on shell in the two-body system
cannot be separated from operators involving more fields,
for example three-body forces of the type 
$(\psi^{\dagger}\psi )^3$. 
Since we have integrated out antiparticles, operators with six or more
fields do not contribute to the two-body system.

For simplicity we work in the center-of-mass frame,
where the scattering particles have relative momenta 
$\vec{p}$ and $\vec{p}\,\rq{}$ before and after scattering, respectively.
From the Feynman diagrams corresponding to the Lagrangian \eqref{eq.2}
we can obtain the nonrelativistic potential in momentum 
space \cite{Phillips:1997xu,Beane:1997pk,vanKolck:1998bw},
\begin{equation}
\bra{\vec{p}\,\rq}\hat{V}\ket{\vec{p}\,} =
\frac{4\pi}{m}\left[
C_{0} 
+ \frac{C_{2}}{2}\left(\vec{p}\,\rq{}^{2} + \vec{p}^{\,2}\right)
+ \frac{C_{4}}{4}\left(\vec{p}\,\rq{}^{2} + \vec{p}^{\,2}\right)^2
+ \ldots 
\right]
\,,
\label{eq.3}
\end{equation}
and, from that, the potential in coordinate space,
\begin{eqnarray}
\bra{\vec{r}\,\rq}\hat{V}\ket{\vec{r}\,} &=& 
\frac{4\pi C_{0}}{m} \, \delta(\vec{r}\,\rq) \delta(\vec{r}) 
- \frac{2\pi C_{2}}{m}
\left[\left(\nabla\rq^2 \delta(\vec{r}\,\rq)\right) \delta(\vec{r}) 
+ \delta(\vec{r}\,\rq) \left(\nabla^2\delta(\vec{r})\right) \right]
\nonumber\\
&&+ \frac{\pi C_{4}}{m}\left[
\left(\nabla\rq^4 \delta(\vec{r}\,\rq)\right) \delta(\vec{r})
+ 2\left(\nabla\rq^2 \delta(\vec{r}\,\rq)\right)
\left(\nabla^2\delta(\vec{r})\right)
+ \delta(\vec{r}\,\rq)\left(\nabla^4\delta(\vec{r})\right)\right] 
+ \ldots \, .
\label{eq.4}
\end{eqnarray}

Our first task is to order these interactions 
according to their effects on observables.
In general the same operator contributes to various orders,
so we decompose the LECs as
\begin{equation}
C_{n} = C_{n}^{(0)} + C_{n}^{(1)} + \ldots \,,
\label{eq.5}
\end{equation}
where $C_{n}^{(N)}$ is the part of $C_{n}$ that contributes at
order $N$. 

Since we are interested in situations where there are
two low-energy $S$-wave poles, the denominator
of the LO scattering amplitude must be quadratic in momentum. 
In our EFT with momentum-dependent interactions we need to treat both $C_0$ 
and $C_2$ nonperturbatively ({\it i.e.} at LO) to reproduce this non-analytic 
behavior. 
In momentum space, the LO potential is therefore
\begin{equation}
\bra{\vec{p}\,\rq}\hat{V}^{(0)}\ket{\vec{p}\,} =
\frac{4\pi}{m}\left[C_{0}^{(0)}
+ \frac{C_{2}^{(0)}}{2}\left(\vec{p}\,\rq{}^{2} + \vec{p}^{\,2}\right) 
\right] \, .
\label{eq.6}
\end{equation}
In other words, $C_{n\ge 4}^{(0)}=0$.
We seek an exact, nonperturbative solution of this potential
under a momentum cut-off regulator $\Lambda$,
with $C_{0,2}^{(0)}(\Lambda)$ determined from the requirement
that two observable quantities be reproduced.

At higher orders, we expect small, perturbative corrections
from higher-derivative operators. 
The fact that zero- and two-derivative operators are
taken as LO might suggest that there is no expansion that suppresses the
higher-derivative operators. 
However, as discussed in the introduction,
we are facing a fine-tuned situation where a low-energy scale
$M_{lo}$ enhances all operators, but not in the same way.
Subleading interactions are suppressed by
powers of the high-energy scale $M_{hi}$,
which are not entirely obvious at first.
A powerful guide in this situation is the renormalization group (RG).
At any given order, observables not used in the determination of 
LECs have residual cutoff dependence: they depend on $M_{lo}$ and $M_{hi}$
in a form dictated by the explicit solution of the EFT interactions
up to that order, and additionally on inverse powers of
$\Lambda$. This residual cutoff dependence will be removed at higher
orders by other LECs,
which will have bare components that scale as inverse powers of $\Lambda$.
Naturalness assumes that the magnitude of a renormalized
LEC is determined by changes in the cutoff of relative 
${\cal O}(1)$. This implies that for cutoffs that do not intrude
in the region where we want the EFT to converge, $\Lambda\simge M_{hi}$,
we can determine the magnitude of the renormalized LEC,
and hence its order, by the replacement $\Lambda\to M_{hi}$ in the bare LEC.

We will show below that the assumption of naturalness for
subleading operators leads to a controlled expansion.
In particular, the NLO momentum-space potential is
\begin{equation}
\bra{\vec{p}\,\rq}\hat{V}^{(1)}\ket{\vec{p}\,} =
\frac{4\pi}{m}\left[C_{0}^{(1)}
+ \frac{C_{2}^{(1)}}{2}\left(\vec{p}\,\rq{}^{2} + \vec{p}^{\,2}\right)
+ \frac{C_{4}^{(1)}}{4}\left(\vec{p}\,\rq{}^{2} + \vec{p}^{\,2}\right)^2
\right]
\,,
\label{eq.7}
\end{equation}
which is to be treated in first-order perturbation theory.
The appearance of quartic momentum operators at this order
means that an additional observable is needed to fix $C_{4}^{(1)}(\Lambda)$.
The latter induces changes in the observables fitted 
at LO. To compensate, we include perturbative changes in $C_{0,2}$
so that $C_{0,2}^{(1)}(\Lambda)$ produce opposite changes in these
observables, which then remain fixed at the values chosen at LO.
For the remaining interactions, $C_{n\ge 6}^{(1)}=0$.

An analogous procedure is followed at higher orders.
The removal of regularization dependence --- up to effects
no larger than those of the truncation in the potential ---
can be performed
by a finite number of parameters {\it at each order}. 
That means that
the theory is renormalizable in the modern sense,
which generalizes the old-fashioned concept
of a finite set of interactions at all orders.

\section{Solving the Schr\"odinger Equation}
\label{seE}

Equipped with the potential we now obtain
the $S$-wave scattering amplitude in an expansion
\begin{equation}
T_{0}(k) = T^{(0)}_{0}(k) + T^{(1)}_{0}(k) + \ldots \, ,
\label{eq.8}
\end{equation}
where successive terms are suppressed by an additional power of $M_{lo}/M_{hi}$.
In order to renormalize the resulting amplitude, we demand that
the $C_n^{(N)}(\Lambda)$ appearing in the expansion
\begin{equation}
\frac{1}{T_{0}(k)} =\frac{1}{T^{(0)}_{0}(k)} 
\left(1- \frac{T^{(1)}_{0}(k)}{T^{(0)}_{0}(k)} + \ldots\right) 
\label{eq.9}
\end{equation}
reproduce the ERE parameters $a_0$, $r_0$, $P_0$, {\it etc.} in 
the inverse of Eq. \eqref{eq.1}.
Then other observables, such as the pole positions, can be predicted.
Other renormalization conditions can be imposed instead.
For example, one can fit the pole positions, if known, at LO 
and predict ERE parameters.
As long as only low-energy input is used, different renormalization
conditions differ only by higher-order effects.

We can obtain the expansion of $T_{0}(k)$
using Feynman diagrams (see App. \ref{AppxA}).
In this case, we have to calculate loop diagrams
which involve the Schr\"odinger propagator.
After renormalization, when positive powers of $\Lambda$
have been removed from loops, each loop effectively contributes
${\cal O}(mk/4\pi)$, while interaction vertices contribute
$4\pi C_nk^{2n}/m$.
The LO potential has to be iterated to all orders
if $C_{n}^{(0)}={\cal O}(1/M_{lo}^{n+1})$,
since then diagrams involving these vertices form a series in $k/M_{lo}$
which, once resummed, can give rise to a pole with $|k|\sim M_{lo}$.
The resummation of Feynman diagrams is equivalent to the
exact solution of the Schr\"odinger equation.
By definition, higher-order LECs have additional inverse powers of $M_{hi}$,
$C_n^{(N)}={\cal O}(1/M_{lo}^{n-N+1}M_{hi}^{N})$.
Subleading interactions are dealt with in distorted-wave perturbation
theory, which is equivalent to a finite number of insertions of
subleading vertices in Feynman diagrams that include all possible LO vertices.
We find the Schr\"odinger equation easier to implement, especially at 
subleading orders,
and we present explicitly here the calculation of $T^{(0,1)}_{0}(k)$.
 
\subsection{Leading order}

If the incoming free-particle state with energy $E\equiv k^2/m$ is denoted 
by $\ket{\vec{k}}$, the scattering amplitude 
from the LO potential $V^{(0)}$, Eq.~\eqref{eq.6}, is 
(see, for example, Ref. \cite{TaylorScattering})
\bea
T^{(0)}_{0} (k) &=& \langle\psi^{(0)}_- | \hat{V}^{(0)} | \vec{k}\rangle 
= \frac{4\pi}{m}\left[\left( C_{0}^{(0)} 
+ \frac{C_{2}^{(0)}}{2} k^2\right){\psi^{(0)}_-}^{*}(0) 
- \frac{C_{2}^{(0)}}{2}{{\psi^{(0)}_-}''}^{*}(0) \right]
\nonumber\\
& = & \langle\vec{k} | \hat{V}^{(0)} | \psi^{(0)}_{+} \rangle 
= \frac{4\pi}{m}\left[\left( C_{0}^{(0)} 
+\frac{C_{2}^{(0)}}{2} k^2\right){\psi^{(0)}_{+}}(0) 
- \frac{C_{2}^{(0)}}{2}{{\psi^{(0)}_{+}}''}(0) \right]
\, ,
\label{eq.10}
\eea
where $|\psi^{(0)}_{+} \,\rangle$ ($|\psi^{(0)}_{-} \,\rangle$) 
is the incoming (outgoing) scattering wavefunction
and ${\psi}''\equiv \nabla^2\psi$.
The scattering wavefunctions are combinations of homogeneous (free-field) and 
particular (potential term with free-field Green's function) solutions
of the Schr\"odinger equation,
\begin{equation}
\psi^{(0)}_{\pm}(\vec{r}) = e^{i\vec{k}\cdot\vec{r}} 
-4\pi \int\!\!\frac{d^3q}{(2\pi)^3}\,
\frac{e^{i\vec{q}\cdot\vec{r}}}{q^2 - k^2 \mp i\epsilon}
\left[\left( C_{0}^{(0)} + \frac{C_{2}^{(0)}}{2}q^2\right)\psi^{(0)}_{\pm}(0)
- \frac{C_{2}^{(0)}}{2}{\psi^{(0)}_{\pm}}''(0)\right]\,.
\label{eq.11}
\end{equation}

We define the integrals
\begin{equation}
I_{2n}^{\pm}(k) = - 4\pi\int\!\!\frac{d^3q}{(2 \pi)^3}\,
\frac{q^{2n}}{q^2 - k^2 \mp i \epsilon} 
= - \sum\limits_{\ell=0}^{\infty}\,L_{1+2(n-\ell)} \, k^{2\ell} \mp  i k^{2n+1}\,, 
\label{eq.12}
\end{equation}
where 
\begin{equation}
L_{\ell} 
=  \theta_{\ell}\Lambda^{\ell} \, ,
\label{eq.13}
\end{equation}
with $\theta_{\ell}$ a regulator-dependent number --- for example, 
$\theta_{\ell}=2/(\ell\pi)$ for a sharp-cutoff regulator.
These integrals allow us to write
\bea
\psi^{(0)}_{\pm}(0) & = & 1 
- \left(C_{0}^{(0)}\, \psi^{(0)}_{\pm}(0)
- \frac{C_{2}^{(0)}}{2} \,{\psi^{(0)}_{\pm}}''(0)\right)I_{0}^{\pm}(k) 
- \frac{C_{2}^{(0)}}{2} \,\psi^{(0)}_{\pm}(0) \, I_{2}^{\pm}(k)
\, , 
\label{eq.14}\\
{\psi^{(0)}_{\pm}}''(0) & = & - k^2 
+ \left( C_{0}^{(0)}\, \psi^{(0)}_{\pm}(0)
- \frac{C_{2}^{(0)}}{2} \,{\psi^{(0)}_{\pm}}''(0)\right) I_{2}^{\pm}(k) 
- \frac{C_{2}^{(0)}}{2} \,\psi^{(0)}_{\pm}(0) \, I_{4}^{\pm}(k) 
\, . 
\label{eq.15}
\eea
Using Eqs.~\eqref{eq.12}, \eqref{eq.14} and \eqref{eq.15} we solve for 
$\psi^{(0)}_{\pm}(0)$ and ${\psi^{(0)}_{\pm}}''(0)$ 
in terms of $C_{0,2}^{(0)}$, $L_{5,3}(k)$ and $I_{0}^{+}(k)$.
We have 
\begin{equation}
\frac{4\pi}{m\, T^{(0)}_{0}(k)} =   
\frac{1}{v_0^{(0)} + v_2^{(0)}k^2} - I^+_{0}(k) \,,
\label{eq.16}
\end{equation}
where
\bea
v_0^{(0)}&=& \frac{C_0^{(0)}-C_2^{(0)2}L_5/4}
{(1+C_{2}^{(0)}L_{3}/2)^2} \,, 
\label{eq.17}
\\
v_2^{(0)}&=& C_2^{(0)}\frac{1+C_2^{(0)}L_3/4}
{(1+C_{2}^{(0)}L_{3}/2)^2} \,,
\label{eq.18}
\eea
a result obtained in Refs. \cite{Phillips:1997xu,Beane:1997pk}.
Equation \eqref{eq.16} is the same $T$ matrix we would have obtained
with an energy-dependent potential $v_0^{(0)} + v_2^{(0)}k^2$.
With our momentum-dependent interactions, however,
$v_{0,2}^{(0)}$ differ from $C_{0,2}^{(0)}$ by cutoff-dependent
factors that trace back to the increased singularity of the 
two-derivative term.

As it stands, the $T$ matrix \eqref{eq.16} depends on the regulator
through $I^+_{0}(k)$, $L_{3,5}$, and $C_{0,2}^{(0)}$.
In particular, the linear cutoff dependence of $I^+_{0}(k)$ needs 
to be eliminated.
Our renormalization procedure involves expanding 
Eq.~\eqref{eq.16} in powers of $k/\Lambda$ 
and equating it to the ERE in Eq.~\eqref{eq.1}.
The unitarity term $ik$ stems from $I^+_{0}(k)$, and the
first two powers of $k^{2}$ 
allow us to express the bare parameters 
$C_{0}^{(0)}(\Lambda)$ and $C_{2}^{(0)}(\Lambda)$ 
as functions of the $L_{n}$ and the observables $a_0$ and $r_{0}$. 
Using Eq. \eqref{eq.13} (details can be found in App. \ref{AppxB}),
we obtain for $r_{0}\ne 0$
\bea
C_{0}^{(0)}(\Lambda) & = &
\frac{\theta_{5}}{\theta_{3}^{2} \Lambda}
\left[1
\mp 2\varepsilon
+\left(1-\frac{\theta_{3}^{2}}{\theta_{5}\theta_{1}}\right)\varepsilon^2 
\pm \left(1- \frac{\theta_{3}\theta_{-1}}{\theta_{1}^{2}} - \frac{\theta_{3} r_0}
{\theta_{1}^{3} a_{0}}\right)\varepsilon^3
+ \mathcal{O}\left(\varepsilon^4, \frac{r_0}{a_0}\varepsilon^4\right) 
\right], 
\label{eq.19} \\ 
\notag \\
C_{2}^{(0)}(\Lambda)  & = &-\frac{2}{\theta_{3}\Lambda^3} 
\left[1 
\mp \varepsilon
\pm \left(1- \frac{\theta_{3}\theta_{-1}}{\theta_{1}^{2}} 
- \frac{\theta_{3} r_0}{\theta_{1}^{3} a_{0}}\right)\frac{\varepsilon^3}{2}
+ \mathcal{O}\left(\varepsilon^5, \frac{r_0}{a_0}\varepsilon^5\right) \right],
\label{eq.20}
\eea
where we introduced
\begin{equation}
\varepsilon = \left(-\frac{2 \theta_{1}^{2}}{\theta_{3}r_{0}\Lambda}\right)^{1/2}\,.
\label{eq.21}
\end{equation}
There are two solutions indicated by the $\pm$ signs
accompanying the unusual inverse powers of $(- r_{0} \Lambda)^{1/2}$.
For real $C_{0}^{(0)}$ and $C_{2}^{(0)}$,
we see from Eqs.~\eqref{eq.19} and \eqref{eq.20} that we must have $r_{0} < 0$.
This result is consistent with the Wigner bound
\cite{Wigner:1955zz}, which puts a condition on the rate of 
change of the phase shift with respect to the energy for a finite-range,
energy-independent potential.
It translates \cite{Fewster:1994sd,Phillips:1996ae}
into a constraint on the effective range $r_{0}$,
\bea
r_{0} \leq 2R \left(1 - \frac{R}{a_{0}} + \frac{R^2}{3a_{0}^2}\right)\, .
\label{eq.22}
\eea
Interpreting the range $R$ of the potential as the 
inverse cutoff in momentum space, 
$R \sim 1/{\Lambda}$, the limit $\Lambda\to\infty$ 
corresponds to a zero-range interaction ($R\to 0$).
In this limit 
there is no positive effective range consistent with Eq.~\eqref{eq.22}
\cite{Beane:1997pk}. 
The requirement of renormalizability for our LO interactions
automatically yields the same constraint.

Now we substitute the expressions for $C_{0}^{(0)}$ and $C_{2}^{(0)}$ 
in terms of the scattering length, the effective range and the cutoff 
into the expression \eqref{eq.16}
and expand $T^{(0)}_{0} (k)$ about $k/\Lambda\to 0$,
\bea
T^{(0)}_{0}(k) &=& 
-\frac{4\pi}{m}
 \left(-\frac{1}{a_{0}} + \frac{r_{0}}{2}k^2 -ik\right)^{-1} 
\left[1
+\left(-\frac{1}{a_{0}} + \frac{r_{0}}{2}k^2 -ik\right)^{-1} 
\frac{r_{0}^2}{4\theta_{1}\Lambda}k^4 +\dots \right]
\label{eq.23} \\
& = & 
-\frac{4\pi}{m}
 \left(-\frac{1}{a_{0}} + \frac{r_{0}}{2}k^2 -ik
 +\frac{r_{0}^2}{4\theta_{1}\Lambda}k^4\right)^{-1} 
+\ldots \,.
\label{eq.24}
\eea
Because the renormalized $C_{0}^{(0)}$ and $C_{2}^{(0)}$ involve only $M_{lo}$,
the ERE parameters scale as $a_{0}\sim r_{0}\sim {1}/{M_{lo}}$,
as needed for a shallow resonance.
For $k\sim M_{lo}$, the first three terms in the denominator on the right-hand 
side of Eq. \eqref{eq.24} are $\mathcal{O}(M_{lo})$,
while the fourth term can be made arbitrarily small as $\Lambda$ increases.
The first three terms generate two poles, including a resonance,
which we discuss in Sec. \ref{poles}, and give rise to the LO phase shift
\begin{equation}
k\cot \delta_0^{(0)}(k) = -\frac{1}{a_{0}} + \frac{r_{0}}{2}\, k^2
 \,.
\label{eq.25}
\end{equation}
For $\Lambda \simge  M_{hi}$, 
the fourth term is no larger than $\mathcal{O}({M_{lo}^2}/{M_{hi}})$.
Barring further fine tuning, this term is comparable to NLO interactions
which will remove its residual cutoff dependence.
In fact, it can be thought as an induced shape parameter
$1/(4\theta_{1}r_0\Lambda)$.
Taking $\Lambda \sim M_{hi}$ in 
\begin{equation}
\Delta\left(k\cot \delta_0^{(0)}(k)\right) = 
\frac{r_{0}^2}{4\theta_{1}\Lambda} \, k^4 
\label{eq.26}
\end{equation}
gives an estimate of the error in $k\cot \delta_0^{(0)}(k)$.
In this case, the right-hand side of Eq. \eqref{eq.26} is indeed
${\cal O}(M_{lo}/M_{hi})$ relative to Eq. \eqref{eq.25}.

At this point we have shown, following
Refs. \cite{Phillips:1997xu,Beane:1997pk},
that the EFT can produce the first two terms in the ERE at LO,
as long as the effective range $r_0<0$.
In the next section we show how subleading contributions systematically 
improve the LO result.
 
\subsection{Subleading order}

Adding perturbations to the LO implies that the corresponding changes in 
the scattering amplitude should be treated in (distorted-wave)
perturbation theory. 
The first correction to the $T$ matrix, $T^{(1)}_{0}(k)$, is linear in 
the parameters of the NLO potential, 
the $C_{0,2,4}^{(1)}$ defined in Eq.~\eqref{eq.5}.
After renormalization,
$C_{4}^{(1)}={\cal O}(1/M_{lo}^4M_{hi})$ since its contribution
must be comparable
to the LO error in Eq.~\eqref{eq.26}.
And, since they are of the same order, 
$C_{0}^{(1)}={\cal O}(1/M_{hi})$
and $C_{2}^{(1)}={\cal O}(1/M_{lo}^2M_{hi})$.

Explicitly (see Ref. \cite{TaylorScattering} again),
\bea
T^{(1)}_{0} (k) & = & \langle\psi^{(0)}_{-} | V^{(1)} | \psi^{(0)}_{+} \rangle 
= \langle\psi^{(0)}_{+}  | V^{(1)} | \psi^{(0)}_{-} \rangle \notag \\
& = & \frac{4\pi}{m}
\left[
C_{0}^{(1)}\,{\psi^{(0)}_{-}}^{*}(0) \,{\psi^{(0)}_{+}}(0) 
- C_{2}^{(1)}\,{\psi^{(0)}_{-}}^{*}(0) \,{\psi^{(0)}_{+}}''(0) 
+ \frac{C_{4}^{(1)}}{2}\left({\psi^{(0)}_{-}}^{*}(0) \,{\psi^{(0)}_{+}}''''(0) 
+ {{\psi^{(0)}_{-}}^{*}}''(0) \,{\psi^{(0)}_{+}}''(0) \right) 
\right]\, , \qquad
\label{eq.27} 
\eea
where ${\psi^{(0)}_{+}}''''\equiv \nabla^4\psi^{(0)}_{+}$ is given by
\begin{equation}
{\psi^{(0)}_{\pm}}''''(0) = k^4 
- \left(C_{0}^{(0)}\,\psi^{(0)}_{\pm}(0)
- \frac{C_{2}^{(0)}}{2}\,{\psi^{(0)}_{\pm}}''(0)\right) I_{4}^{\pm}(k) 
- \frac{C_{2}^{(0)}}{2}\,\psi^{(0)}_{\pm}(0) \, I_{6}^{\pm}(k) 
\, . 
\label{eq.28}
\end{equation}
Once this is substituted in Eq. \eqref{eq.27},
\bea
\frac{T^{(1)}_0 (k)}{T^{(0)2}_0 (k)} & = & \frac{m}{4\pi}
\frac{v_0^{(1)} + v_2^{(1)} k^2 + v_4^{(1)} k^4}{(v_0^{(0)} + v_2^{(0)} k^2)^2 }
\,,
\label{eq.29}
\eea
where
\bea
v_0^{(1)}&=&  \frac{1}{(1+C_{2}^{(0)} L_{3}/2)^2}
\Biggl\{C_{0}^{(1)} 
- C_{2}^{(1)}
\frac{C_0^{(0)} L_{3} + C_2^{(0)} L_5/2}{1+C_2^{(0)} L_3/2} 
+\frac{C_{4}^{(1)}/2}{(1+C_2^{(0)} L_3/2)^2}
\left[C_0^{(0)} L_3 \left(C_0^{(0)} L_3 + C_2^{(0)} L_5/2 \right)
\right.
\nonumber\\
&& \left.
\qquad\qquad\qquad\qquad
- C_0^{(0)} L_5 
+ C_2^{(0) 2} 
\left(1+C_2^{(0)} L_3/4\right) \left(L_5^2 - L_3 L_7\right)/2 
- C_2^{(0)} L_7/2 \right]
\Biggr\}
\,,
\label{eq.30}
\\
v_2^{(1)}&=&  \frac{1}{(1+C_{2}^{(0)} L_{3}/2)^3}
\left[C_{2}^{(1)}
- \frac{C_{4}^{(1)}}{2}
\left(C_{0}^{(0)}L_{3} + C_{2}^{(0)} L_{5}
+ 2 \frac{C_{0}^{(0)} L_{3} + C_{2}^{(0)} L_{5}/2}{1+C_2^{(0)} L_3/2}
\right)\right]
\,,
\label{eq.31}
\\
v_4^{(1)}&=& C_{4}^{(1)}
\frac{1+C_{2}^{(0)} L_{3}/4}{(1+C_2^{(0)} L_3/2)^4}
\,.
\label{eq.32}
\eea
Again, Eq. \eqref{eq.29} is the result we would obtain by treating an
energy-dependent potential $v_0^{(1)} + v_2^{(1)} k^2 + v_4^{(1)} k^4$
in first-order perturbation theory. 
In our case, however, the additional singularity of two- and four-derivative
interactions leads to the relations 
\eqref{eq.30}, \eqref{eq.31}, and \eqref{eq.32}.

As before, we renormalize the NLO amplitude by expanding the right-hand side
of Eq. \eqref{eq.9} in powers of $k/\Lambda$ and matching it to the ERE,
Eq. \eqref{eq.1}. 
The presence of $C_{4}^{(1)}={\cal O}(1/M_{lo}^4M_{hi})$ 
now ensures that the quartic power
of momentum can be made cutoff-independent by demanding that it
reproduce the empirical value of the shape parameter $P_0$.
The other two NLO parameters, $C_{0,2}^{(1)}$, can be chosen
so that the terms independent of and quadratic in momentum
remain unchanged. Again, with details given in App. \ref{AppxB},
we find
\bea
C_{0}^{(1)}(\Lambda) & = & 
\frac{\theta_{1}^{2}}{2\theta_{3}^{4}}\left(\theta_{7}\theta_{3}-2\theta_{5}^2\right)
P_{0} r_{0} 
\Biggl\{ 1 
- \left[ 2 - \frac{\theta_{5}(\theta_{5}\theta_{1}-2\theta_{3}^{2})}
{\theta_{1}(\theta_{7}\theta_{3}-2\theta_{5}^{2})}\right] \varepsilon
- \left[
\frac{\theta_{5}^{2}\theta_{1}^{2}-3\theta_{5}\theta_{3}^{2}\theta_{1}+\theta_{3}^{4}}
{\theta_{1}^{2}(\theta_{7}\theta_{3}-2\theta_{5}^{2})} 
- \frac{2\theta_{3}\theta_{-1}}{\theta_{1}^{2}}
- \frac{\theta_{3}r_{0}}{\theta_{1}^{3} a_{0}}\right] \varepsilon^2
\notag \\
&& \qquad \qquad \qquad \qquad \qquad 
+ \,\mathcal{O}\left(\varepsilon^3, \frac{r_{0}}{a_{0}}\varepsilon^3\right)
\Biggr\} 
+\frac{\theta_{1}}{\theta_{3}^4\Lambda}
\left(\theta_{7}\theta_{3}-2\theta_{5}^{2}\right)
\left[1+\mathcal{O}\left(\varepsilon\right)\right]\, ,
\label{eq.33} \\ 
C_{2}^{(1)}(\Lambda) & =  & 
\frac{3\theta_{5}\theta_{1}^{2}}{2 \theta_{3}^{3}}\,\frac{P_{0} r_{0}}{ \Lambda^{2}} 
\left[1
- \left(5 - \frac{2\theta_{3}^2}{\theta_{5}\theta_{1}}\right)
\frac{\varepsilon}{3}
- \left(\frac{2}{3} + \frac{\theta_{3}^{2}}{3\theta_{1}\theta_{5}} 
- \frac{2\theta_{3}\theta_{-1}}{\theta_{1}^{2}}
- \frac{\theta_{3} r_{0}}{\theta_{1}^{3} a_{0}}\right)\varepsilon^2
+ \mathcal{O}\left(\varepsilon^3, \frac{r_{0}}{a_{0}}\varepsilon^3\right)\right]
\notag \\
&& 
+\frac{3\theta_{5}\theta_{1}}{\theta_{3}^3\Lambda^3}
\left[1+\mathcal{O}\left(\varepsilon\right) \right]
\, ,
\label{eq.34}\\ 
C_{4}^{(1)}(\Lambda) & = & 
-\frac{\theta_{1}^{2}}{\theta_{3}^{2}} \,\frac{P_{0} r_{0}}{\Lambda^{4}}
\left[1 - \varepsilon
-\left(1 -\frac{2 \theta_{3} \theta_{-1}}{\theta_{1}^{2}}
-\frac{\theta_{3} r_{0}}{\theta_{1}^{3} a_{0}}\right) \varepsilon^2
+ \mathcal{O}\left(\varepsilon^3, \frac{r_{0}}{a_{0}}\varepsilon^3\right)
\right]
-\frac{2\theta_{1}}{\theta_{3}^2\Lambda^5}
\left[1+\mathcal{O}\left(\varepsilon\right) \right]
\, .
\label{eq.35}
\eea
The terms independent of $P_0$ are those induced by the residual
cutoff dependence in the quartic term of Eq. \eqref{eq.24}.
The others are all linear in the NLO physical parameter $P_0$,
whose sign is not constrained by renormalization at this order.
The situation here is analogous to the one-pole case 
\cite{vanKolck:1997ut,Kaplan:1998tg,Kaplan:1998we,vanKolck:1998bw},
where the perturbative treatment of $C_2$ places no
renormalization constraints on the sign of $r_0$.

At this point we have tuned $C_{0}$, $C_{2}$ and $C_{4}$ in such a way that they 
reproduce the 
ERE in Eq.~\eqref{eq.1} exactly in the limit $\Lambda\rightarrow\infty$. 
For a large but finite $\Lambda$, the NLO EFT reproduces  
Eq.~\eqref{eq.1} up to a correction that goes as the sixth power of momentum,
\begin{equation}
T_{0}^{(0+1)}(k) = -\frac{4\pi}{m}
\left[-\frac{1}{a_{0}} + \frac{r_{0}}{2} k^2 
- P_{0}\left(\frac{r_{0}}{2}\right)^3 k^4 - ik
-\frac{P_{0}r_{0}^4}{8\theta_{1}\Lambda} k^6 
\right]^{-1} +\ldots \, . 
\label{eq.36}
\end{equation}
Because $C_{4}={\cal O}(1/M_{lo}^4M_{hi})$,
here the shape parameter $P_{0}={\cal O}(M_{lo}/M_{hi})$.
The phase shift up to NLO is given by
\begin{equation}
k\cot \delta_0^{(0+1)}(k) = -\frac{1}{a_{0}} + \frac{r_{0}}{2}k^2 
- P_{0} \left(\frac{r_{0}}{2}\right)^3 k^4 
\,,
\label{eq.37}
\end{equation}
with an error that can be estimated from the $k^6$ term 
once we take $\Lambda\to M_{hi}$ in
\begin{equation}
\Delta\left(k\cot \delta_0^{(0+1)}(k)\right) = 
\frac{|P_{0}| r_{0}^4}{8 |\theta_{1}| \Lambda} \, k^6 \,.
\label{eq.38}
\end{equation}
This error shows that at N$^2$LO a 
$(\psi^\dagger \psi) (\psi^{\dagger}\overleftrightarrow{\nabla}^6\psi)$ 
interaction with LEC $4\pi C_6/m$ is needed,
where after renormalization $C_{6}={\cal O}(1/M_{lo}^5M_{hi}^2)$.
At this order this interaction is solved in first-order distorted-wave
perturbation theory, while NLO interactions need to be accounted for
in second order.
The procedure continues at higher orders in an obvious way. 

\section{Poles and Residues}
\label{poles}

The distinctive feature of our LO is the existence of two 
$S$-wave $S$-matrix poles,
allowing for the possibility of a resonance.
Our $S$-wave $S$ matrix can be written as 
\begin{equation}
S_{0}(k)=1-\frac{imk}{2\pi}T_{0}(k)
= e^{2i\phi(k)} \frac{(k+k_1)(k+k_2)}{(k-k_1)(k-k_2)}
\, ,
\label{eq.39}
\end{equation}
where $k_{1,2}\ne 0$ denote the pole positions and 
$\phi(k)$ is the nonresonant or ``background'' contribution to the phase shift.
These quantities can be expanded order by order,
and we now obtain them up to NLO.

\subsection{Leading order}

The $S$ matrix corresponding the LO $T$ matrix \eqref{eq.24},
$S_{0}^{(0)}(k)$, is nothing
but that corresponding to the effective-range approximation to the ERE,
that is, Eq. \eqref{eq.39} with \cite{Peierls:1959}
\begin{equation}
k^{(0)}_{1,2} =   \frac{1}{r_{0}} \left(i \pm \sqrt{\frac{2r_{0}}{a_0} - 1} \right)
\label{eq.40}
\end{equation}
and
\begin{equation}
\phi^{(0)}(k)=0 
\, . 
\label{eq.41}
\end{equation}
While the effective range $r_{0}<0$ for renormalization, the scattering length 
$a_{0}$ can be positive or negative, giving rise to five qualitatively distinct 
cases:

\begin{enumerate}
\item $\quad -2|r_{0}|<a_{0}<0$: there are two resonance poles 
\bea
k_{1}^{(0)} &=&  k_{R} - i k_{I} \, , 
\label{eq.42}\\
\qquad
k_{2}^{(0)} &=& - k_{R} - i k_{I} \, ,
\label{eq.43}
\eea
with 
\bea
k_{I} & = & \frac{1}{|r_{0}|} >0 \, ,
\label{eq.44} \\
k_{R} & = & \frac{1}{|r_{0}|}\sqrt{\frac{2r_{0}}{a_0} - 1}>0
\, .
\label{eq.45}
\eea
The two poles are thus forced to be in the lower half of the complex momentum
plane by the requirement of renormalizability.
This is in agreement with the general requirement on the $S$ matrix
\cite{Moller:1946,Hu:1948zz,Schuetzer:1951} 
that leads to states decaying with time. 
In the limit $\Lambda\to \infty$ we obtain the well-known form 
\begin{equation}
S^{(0)}_{0}(E) = \frac{E-E_0^{(0)}-i\Gamma^{(0)}(E)/2}{E-E_0^{(0)}+i\Gamma^{(0)}(E)/2} 
\,,
\label{eq.46}
\end{equation}
where
\bea
E_0^{(0)} &=& \frac{k_{R}^2+k_{I}^2}{m}= \frac{2}{ma_0r_0} >0
\,, \label{eq.47}\\
\Gamma^{(0)}(E) &=&  \frac{4kk_{I}}{m} = \frac{4}{m|r_0|}\sqrt{mE} >0
\,. \label{eq.48}
\eea
The residues of $iS_0^{(0)}$ at the poles are complex,
\begin{equation}
\left. \mathrm{Res}\!\left(iS_0^{(0)}\right)\right|_{\pm k_R-ik_I} =  
2k_{I}\left(1\mp i\frac{k_{I}}{k_{R}}\right) 
\, .
\label{eq.49} 
\end{equation}
Our power counting describes the situation $|a_{0}| \sim |r_{0}| \sim 1/M_{lo}$, 
where the resonance is shallow but broad in the sense that 
$k_{I} \sim k_{R}\sim M_{lo}\ll M_{hi}$. 
We cannot exclude a situation where $r_{0}\gg a_{0}$ and
the resonance is narrow, that is, $k_{R}\gg k_{I}$ with 
nearly real and positive residues \cite{Peierls:1959}. 
However, our power counting is somewhat artificial
in this case, since the unitarity
term $-ik$ in Eq. \eqref{eq.24} is small compared to 
the inverse scattering length
and the effective range, except near the pole.
A narrow resonance arises more naturally from a
dimeron field where residual mass and kinetic terms are treated as LO,
while loops are included perturbatively except in the vicinity of the
resonance \cite{Bedaque:2003wa}.

\item $\quad a_{0}=-2|r_{0}|$: there is a double pole on the negative
imaginary axis \cite{DemkovDrukarev1966},
\begin{equation}
k_{1}^{(0)}= k_{2}^{(0)} \equiv k_{1\equiv 2}^{(0)} =  -\frac{i}{|r_{0}|} \, ,
\label{eq.50}
\end{equation}
with positive residue
\begin{equation}
\left.\mathrm{Res}\!\left(iS_0^{(0)}\right)\right|_{k_{1\equiv 2}} = 4i k_{1\equiv 2}
>0 
\, .
\label{eq.51}
\end{equation}
This represents a virtual state. 
A double pole on the upper plane is again excluded
by the requirement of renormalizability, in agreement
with other arguments \cite{Moller:1946,Hu:1948zz,DemkovDrukarev1966}.

\item $\quad a_{0}<-2|r_{0}|$: there are two virtual states
represented by poles on the negative imaginary axis,
\bea
k_{1}^{(0)} &=&  -i \kappa_{-} \, , 
\label{eq.52}\\
\qquad
k_{2}^{(0)} &=&  -i \kappa_{+} \, ,
\label{eq.53}
\eea
where
\begin{equation}
\kappa_\pm= \frac{1}{|r_{0}|} \left(1\pm \sqrt{1-\frac{2r_{0}}{a_{0}}} \right) >0
\, .
\label{eq.54}
\end{equation}
They have residues of opposite signs,
\begin{equation}
\left.\mathrm{Res}\!\left(iS_0^{(0)}\right)\right|_{-i\kappa_\pm} = 
\pm 2 \kappa_\pm\, \frac{\kappa_{+} + \kappa_{-}}{\kappa_{+} - \kappa_{-}}
\label{eq.55} \,,
\end{equation}
the shallowest pole ($k_1$) with the negative sign.

\item $\quad a_{0}>0$: the two poles are on opposite sides of the imaginary
axis,
\bea
k_{1}^{(0)} &=&  i \kappa_{-} \, , 
\label{eq.56}\\
\qquad
k_{2}^{(0)} &=&  -i \kappa_{+} \, ,
\label{eq.57}
\eea
where
\begin{equation}
\kappa_\pm = \frac{1}{|r_{0}|} \left(\sqrt{1+\frac{2|r_{0}|}{a_{0}}} \pm 1\right)
>0\, .
\label{eq.58}
\end{equation}
Both residues are positive,
\begin{equation}
\left.\mathrm{Res}\!\left(iS_0^{(0)}\right)\right|_{\mp i \kappa_\pm} = 
2 \kappa_\pm\, \frac{\kappa_{+} - \kappa_{-}}{\kappa_{+} + \kappa_{-}}>0
\label{eq.59} \,.
\end{equation}
This indicates that the pole $k_1$ on the positive imaginary axis 
is a bound state \cite{Moller:1946}.
The constraint $r_0<0$ from renormalization thus excludes the possibility
of a ``redundant'' pole \cite{Ma:1946,TerHaar:1946,Ma:1947zz}
on the positive imaginary axis with negative residue. 
If the effective range were positive, such a pole could
arise together with a shallower bound or virtual state for, respectively,
$a_0>2r_0$ or $a_0<0$, see Table \ref{tbl1}. 
The interpretation of
a redundant pole is unclear: it has a non-normalizable wavefunction
\cite{TerHaar:1946,Nelson:1971nr},
but carries information about the asymptotic behavior
of continuum states \cite{Terry:1980bg}.
Since at least in a nonrelativistic setting its position is determined by
the range of the potential \cite{Bargmann:1949zz,Peierls:1959,Yamamoto:1962},
it is comforting that renormalization of our EFT prevents
a shallow redundant pole.

\begin{table}[t]
\begin{center}
\begin{tabular}{|c||c|c|c|c|}
\hline
$a_{0}$, $r_{0}$
 & $-$, $-$ & $+$, $-$ & $-$, $+$ & $+$, $+$\\
$\mathrm{Im}\,k_1$, $\mathrm{Im}\,k_2$ 
 & $-$, $-$ & $+$, $-$ & $-$, $+$ & $+$, $+$\\
$\mathrm{Res}(iS_0^{(0)})|_{k_1}$,
$\mathrm{Res}(iS_0^{(0)})|_{k_2}$ 
 & $-$, $+$ & $+$, $+$ & $-$, $-$ & $+$, $-$\\
pole 1, pole 2 
 & V, V & B, V & V, R & B, R\\
\hline
\end{tabular}
\end{center}
\caption{Character of (simple) poles on the imaginary axis according to
the signs of the scattering length $a_0$ and effective range $r_0$ for
$|\mathrm{Im}\,k_2|>|\mathrm{Im}\,k_1|$.
In the last row V stands for virtual state, B for bound state,
and R for redundant pole.
Only the first two columns are allowed by renormalization
of the EFT.}
\label{tbl1}
\end{table}

\item $\quad |a_{0}|^{-1}= 0$: this is the boundary between the previous two
cases. It is essentially the limit $|a_{0}|\to \infty$ of these cases,
except that the $S$ matrix has a single pole
\begin{equation}
k_{2}^{(0)} =  -i\kappa_+ = -2i |r_0| \, ,
\label{eq.60}
\end{equation}
with positive residue
\begin{equation}
\left.\mathrm{Res}\!\left(iS_0^{(0)}\right)\right|_{-i\kappa_+} = 2 \kappa_+>0
\label{eq.61} \,.
\end{equation}
The other, would-be $S$-matrix pole is only a pole of the $T$ matrix
which is sometimes called a zero-energy resonance.
When $|r_0| \sim 1/M_{hi}$, the corresponding 
EFT \cite{vanKolck:1997ut,Kaplan:1998tg,Kaplan:1998we,vanKolck:1998bw}
is scale invariant at LO with no other low-energy $T$-matrix pole. Here, 
the dimensionful parameter $|r_0|\sim 1/M_{lo}$ explicitly breaks
scale invariance at LO, generating a virtual pole.

\end{enumerate}

When we use effective-range parameters to fix the LECs, 
the pole positions have an uncertainty  $\Delta k^{(0)}$
due to the neglect of higher orders.
We can estimate the magnitude of the error 
from the residual cutoff dependence in 
Eq. \eqref{eq.24} and then varying the cutoff from the theory's
breakdown scale to much larger values. 
We find
\bea
|\Delta k^{(0)}_{1,2}| 
&= &\frac{r_0^2}{4\Lambda} 
\left|\frac{k^{(0)4}_{1,2}}{\theta_{1} \sqrt{\frac{2r_{0}}{a_0} - 1}}\right| 
\, , \quad a_0 \neq 2r_0\, ,
\label{eq.62}\\
&=&\frac{1}{|r_0|}\,\sqrt{\frac{1}{2 |\theta_{1} r_0|\Lambda}}
\, , \quad a_0 = 2r_0\, ,
\label{eq.63}
\eea
with $\Lambda \sim M_{hi}$.
Evidently $\Delta k^{(0)}=0$ if we use the pole positions as input.

\subsection{Subleading order}

If their positions are not used as input, the pole positions will
move slightly at subleading orders and approach, if the theory is converging,
their exact locations. 
As before, the procedure is systematic and we illustrate it only for NLO.

When both poles are simple, that is, for $a_0 \neq 2\,r_0$,
the pole positions can be written as 
\begin{equation}
k^{(0+1)}_{1,2} =   k^{(0)}_{1,2} + k^{(1)}_{1,2} 
\label{eq.64}
\end{equation}
in terms of the NLO shift 
\begin{equation}
k^{(1)}_{1,2} 
=\frac{P_0}{|r_{0}|}\left[
\mp\frac{1}{\sqrt{\frac{2r_{0}}{a_0} - 1}}
\left(\frac{r_0^2}{2a_0^2}-\frac{2r_0}{a_0}+1\right)
+ i \left(1-\frac{r_0}{a_0}\right)
\right]
\, .
\label{eq.65} 
\end{equation}
This is precisely what is needed to cancel 
the spurious double pole in $T_{0}^{(1)}$ 
(the ``good-fit condition'' of Ref. \cite{Mehen:1998zz}).
Now the error is reduced to the size of N$^2$LO interactions when 
$\Lambda$ is above the breakdown scale of the theory,
\begin{equation}
|\Delta k^{(1)}_{1,2}| =  \frac{r_{0}^4 }{8 \Lambda}
\left|\frac{P_0 \, k^{(0)6}_{1,2}}{\theta_{1} \sqrt{\frac{2r_{0}}{a_0}-1}} \right|
\,.
\label{eq.66}
\end{equation}

In contrast, when $a_0 = 2r_0$ the double nature of the pole leads to an 
expansion in half powers for its position,
\begin{equation}
k^{(0+1)}_{1,2} = k^{(0)}_{1\equiv 2} + k^{(1/2)}_{1, 2} + k^{(1)}_{1\equiv 2} 
\,,
\label{eq.67} 
\end{equation}
where
\bea
k^{(1/2)}_{1,2} & = & \pm\,\frac{\sqrt{P_0}}{2 |r_0|}  
\, ,
\label{eq.68} \\
k^{(1)}_{1\equiv 2} & = & i \, \frac{P_0}{2 |r_0|} 
\, .
\label{eq.69} 
\eea
We will refer to the half-power correction as N$^{1/2}$LO.
While the smaller NLO correction is imaginary,
the N$^{1/2}$LO correction is real for $P_0>0$,
indicating that in the underlying theory the resonant poles
have not coalesced. 
Conversely, $P_0<0$ would imply the underlying poles are already
two separated virtual states. 
There is no separate $\Lambda$-dependent error estimate for the N$^{1/2}$LO 
displacement. 
As an estimate for the error we take
\begin{equation}
|\Delta k^{(1/2)}_{1,2}| = \frac{|P_0|}{2 |r_0|}
\,,
\label{eq.70}
\end{equation}
which is $\sqrt{P_{0}} \sim \mathcal{O}((M_{lo}/M_{hi})^{1/2})$ smaller than 
$k^{(1/2)}_{1,2}$ in Eq.~\eqref{eq.68}. This estimate could
easily be off by a factor of $\mathcal{O}(1)$, but it accidentally
exactly coincides with the magnitude of the NLO shift \eqref{eq.69}. 
The $\Lambda$-dependent error of the latter scales as the 3/2 power 
of the expansion parameter,
\begin{equation}
|\Delta k^{(1)}_{1,2}| = \frac{\sqrt{|P_0|}}{4 r_0^2 |\theta_{1}| \Lambda}
\,.
\label{eq.71}
\end{equation}

In either case,
up to higher-order terms the NLO $S$ matrix $S_{0}^{(0+1)}(k)$ is given by 
Eq. \eqref{eq.39} but with the poles at their NLO positions $k^{(0+1)}_{1,2}$ and 
\begin{equation}
\phi^{(0+1)}(k)=\frac{P_0 r_0}{2} \,k \equiv - c \,k
\, .
\label{eq.72}
\end{equation}
The nonresonant contribution to the phase shift is a subleading effect
linear in $k$. This form of the $S$ matrix for short-range forces with 
two poles has been arrived at by causality-type 
arguments \cite{Hu:1948zz,vanKampen:1953}, with $c\ge 0$
related to the range $R$ of the force.
For $c\ge 0$ we have $P_0\ge 0$, although this constraint does not
follow from renormalization of our EFT to NLO.

As we have seen, the renormalization condition on the EFT allows only the 
standard
cases of a (decaying) resonance, a bound state, and virtual states.
The pole positions can be determined from the effective-range 
parameters with increasing precision as the order increases.
In order to show explicitly that the accuracy also improves,
we consider an explicit example of underlying theory next.

\section{Toy Model}
\label{toy}

The inclusion of all interactions allowed by symmetries ensures
that any underlying dynamics producing the same low-energy
pole structure can be accommodated in the EFT.
Information about the
dynamics at short-distance scales is encoded 
in the LECs.
We now consider a simple model for the short-distance physics,
in order to illustrate how the EFT captures the long-distance dynamics
associated to the existence of two shallow poles.

As a toy model we take a potential consisting of an attractive spherical well 
of range $R$ and depth $\beta^2/mR^2$ with a repulsive delta shell with 
strength $\alpha/mR$ at its edge:
\bea
V(r) = \frac{\alpha}{mR}\,\delta(r-R) - \frac{\beta^2}{mR^2}\,\theta(R-r) \,,
\label{eq.73}
\eea
with $\alpha>0$ and $\beta>0$. 
This model was used for resonances in Ref. \cite{Gelman:2009be}.
The Schr\"odinger equation
is easily solved in the $S$-wave in the standard fashion
inside $(r<R)$ and outside $(r>R)$,
with the wavefunctions and their derivatives matched at the range $R$.
One finds 
\begin{equation}
\cot\delta_0(k) = - \frac{\kappa R \cot(\kappa R)\cot(kR) + \alpha\cot(kR)+kR}
{\kappa R \cot(\kappa R) + \alpha -kR \cot(kR)}
\, ,
\label{eq.74}
\end{equation}
where $\kappa=\sqrt{k^2+\beta^2/R^2}$.
Equivalently, the phase shift is given by
\begin{equation}
\delta_0(k)=-k R
+\arctan\!\left(\frac{kR}{\kappa R\,\cot(\kappa R)+\alpha}\right)\,.
\label{eq.75}
\end{equation}

\subsection{Scattering length, effective range, and pole positions}
\label{basic}

We are interested in the dynamics for $k\ll R^{-1}\equiv M_{hi}$.
Equations \eqref{eq.74} and \eqref{eq.75}
reduce to the well-known expressions for the attractive 
spherical well \cite{Moller:1946} when $\alpha=0$. 
In this case there is a single
low-energy pole, either a bound or a virtual state, which is captured
by the EFT where LO consists of only $C_0$ and NLO of $C_2$ 
\cite{vanKolck:1997ut,Kaplan:1998tg,Kaplan:1998we,vanKolck:1998bw}.
For $\alpha>0$, we can have two low-energy poles.
Tuning $\alpha$ and $\beta$ yields the various cases 
(bound, virtual, resonant) considered above.  

\begin{figure}[t]
  \centering
  \includegraphics[scale=0.2]{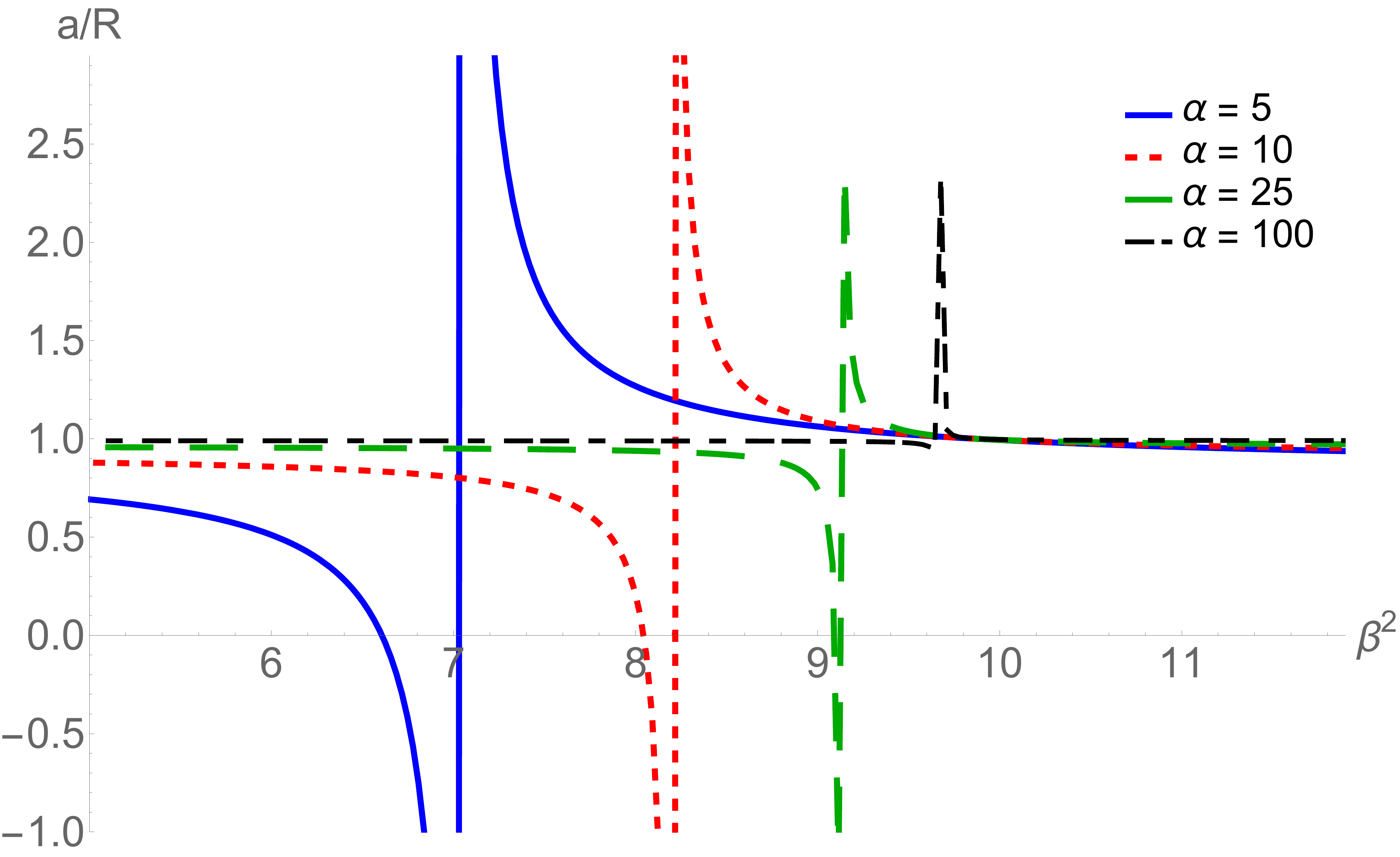}
  \includegraphics[scale=0.2]{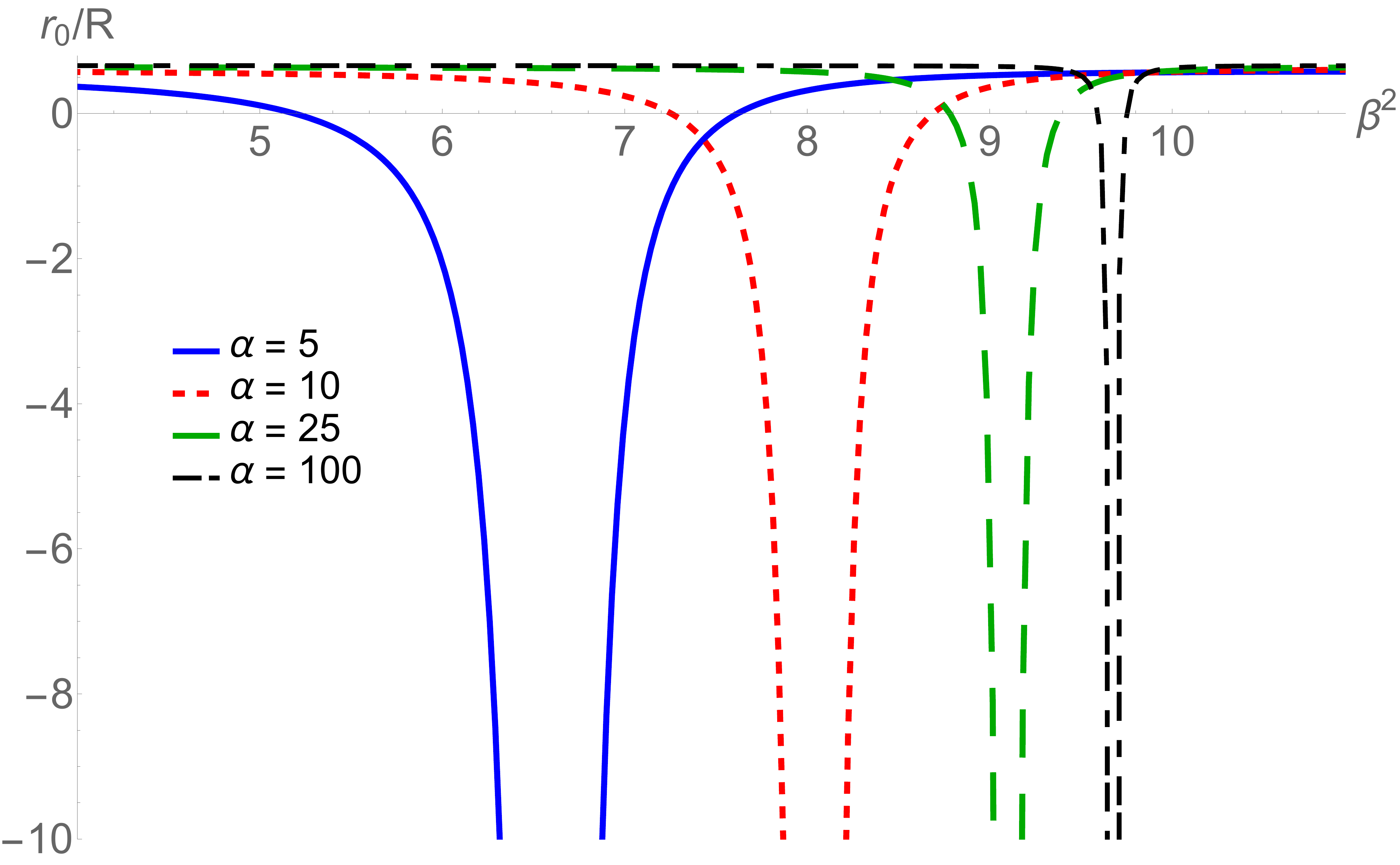}
\caption{Scattering length $a_0$ (left panel) and effective range $r_{0}$
(right panel) in units of the potential range $R$ as functions
of the square of the strength $\beta$ of the spherical well,
for various values of the strength $\alpha$ of the delta shell.}
\label{fig1}
\end{figure}

Expanding the inverse of the $T$ matrix in $kR\ll 1$ and matching to the
ERE,
\bea
\frac{a_{0}}{R} &=& 1-\frac{1}{\alpha+\beta\cot\beta}
\,, \label{eq.76} \\
\frac{r_{0}}{R} &=& \frac{2(\alpha+\beta\cot\beta)\,(\alpha+\beta\cot\beta -3)
-3[\cot^2\beta -(\cot\beta)/\beta -1]}
{3(\alpha +\beta\cot\beta - 1)^2} \label{eq.77}\\
&=& 1 -\frac{1}{3}\left(\frac{R}{a_0}\right)^2
-\frac{R}{\beta^2a_0}
-\frac{\alpha}{\beta^2}
\left[\alpha+1
-2\alpha\frac{R}{a_0}
+ \left(\alpha-1\right)\left(\frac{R}{a_0}\right)^2
\right] 
\, .\label{eq.78}
\eea
The next term in the expansion of $T_{0}^{-1}(k)$ gives an expression for the 
shape parameter $P_{0}$ as a function of the parameters $\alpha$ and $\beta$. 
We plot $a_{0}$ and $r_{0}$ as functions of $\beta$ for various values of 
$\alpha$ in Fig.~\ref{fig1}. 
For most values of the potential parameters, $a_{0}\approx R$ and 
$r_{0}\approx R$ as expected from dimensional analysis.
Only some specific regions have unnaturally large magnitudes.
In the regions where $|a_{0}|\gg R$, $a_0$ can be positive or negative, but it is
still very likely that $r_{0}\approx R$.
This is the situation previously investigated in EFT where only
$C_0$ is enhanced.
Only in narrow parameter ranges where $\beta\cot\beta\approx -\alpha \ll -1$
can we
have both $|a_{0}|\gg R$ and $|r_{0}| \gg R$.
For a pure spherical well ($\alpha=0$), $r_{0}/R$ is given by the first
three terms in Eq. \eqref{eq.78} and it is easy to see that 
$|a_{0}|\gg R$ leads to $r_{0}\approx R$. 
The additional parameter $\alpha>0$ allows $|r_{0}| \gg R$.
But in this case $r_0<0$, just as obtained from renormalization of the EFT
considered here, where also $C_2$ is enhanced.

\begin{figure}[t]
  \includegraphics[scale=0.3]{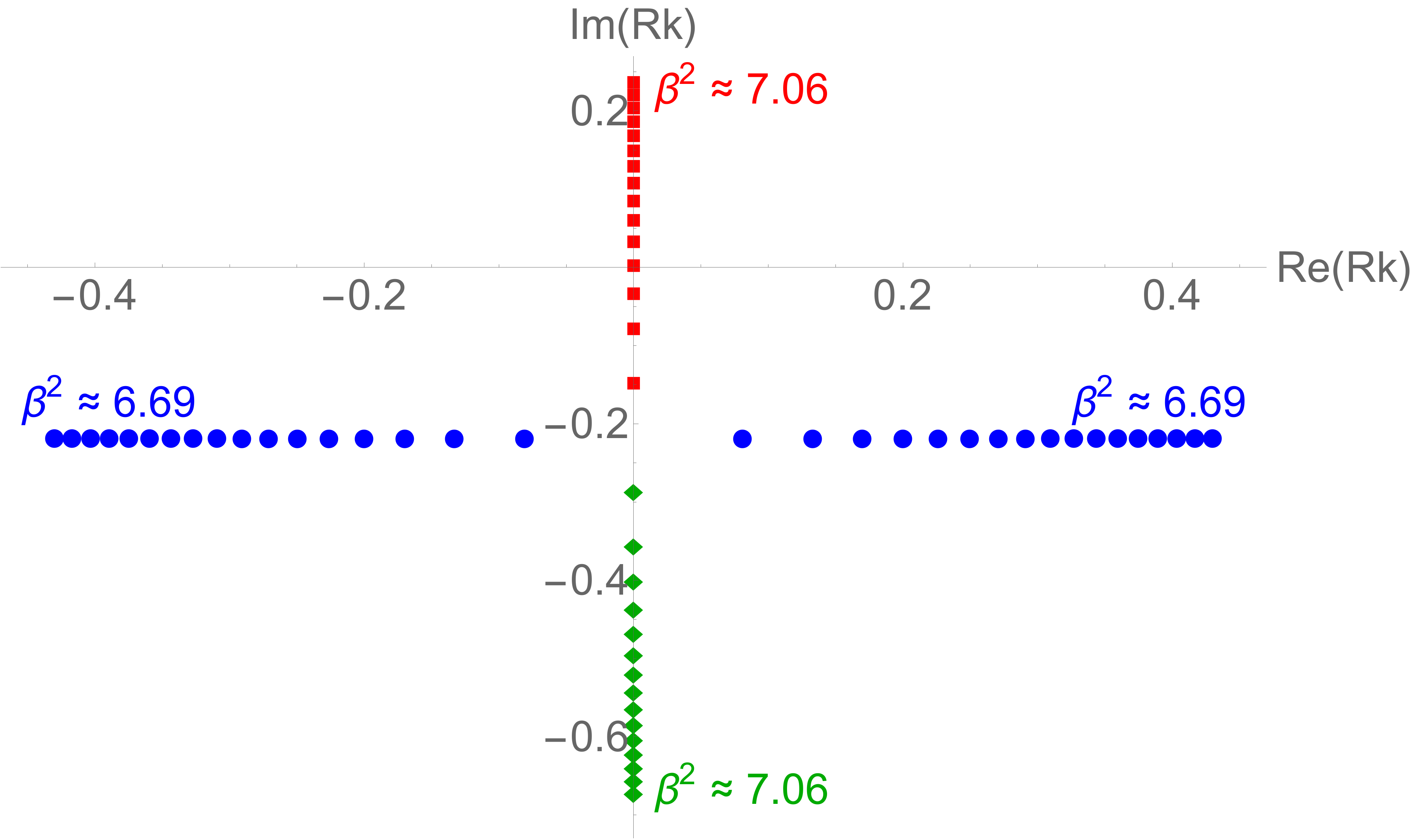}
\caption{Pole positions in the $kR$ complex plane for 
a given strength of the delta-shell potential, $\alpha = 4.68144$.
As the depth $\beta^2$ of the spherical well
increases from 6.69 to 7.06, the two resonance poles
(blue circles) approach the imaginary axis 
and then turn into two virtual poles:
one that remains a virtual state (green diamonds), another that
eventually becomes a bound state (red squares).}
\label{fig2}
\end{figure}

For given values of $a_{0}$ and $r_{0}$, we can solve Eqs.~\eqref{eq.76} and 
\eqref{eq.77} for $\alpha$ and $\beta$, and then use the exact expression for 
$k\cot\delta_0$ in Eq.~\eqref{eq.74} to find the poles of $T_{0}$.
In the regions where $|a_{0}|/R\gg 1$ and $|r_{0}|/R \gg 1$ there are
two low-energy poles. 
An example is shown in Fig.~\ref{fig2}
where we hold the strength of the delta shell fixed at $\alpha = 4.68144$
and slightly vary the depth of the well around $\beta^2\approx 7$. 
In the plot we begin with $\beta^2=6.69$, which gives us two resonance poles,
and increase $\beta^2$, which corresponds to increasing the attraction of the 
well. The two poles collide on the negative imaginary axis creating
a double pole, 
and then one pole moves down as a virtual state
while the other moves up until it eventually becomes a bound state.
We continue following the increase in binding till $\beta^2=7.06$.
The pole evolution covers the situations we considered in the 
previous section.
Note that a similar pole evolution exists for the spherical well
alone ($\alpha=0$) \cite{Nussenzveig:1959}, but in the latter case
the coalescence on the imaginary axis happens at $k_{1}=k_{2}=-i/R$,
which is outside the EFT range. 
The possibility of this evolution for a general potential well surrounded
by a barrier was studied in Ref. \cite{DemkovDrukarev1966}.
We now turn to a quantitative comparison with the EFT.

\subsection{Comparison with the EFT}
\label{sec5}

We now compare the predictions of the EFT at LO and NLO,
using a sharp-cutoff regulator, with the toy model.
We choose values for $a_0$ and $r_0$, from which we extract 
$\beta^2$ and $\alpha$, and calculate $P_0$.
We fit the EFT to these ERE parameters and compare the resulting phase shifts
and pole positions. 
Since the phase shift is very sensitive to changes in momentum $k$ and there 
are periodic discontinuities in its derivatives, it is better to work with 
$k\cot\delta_0$ instead of $\delta_0$ itself.
For the toy model we evaluate $k\cot\delta_0$ from Eq.~\eqref{eq.74}. 
For the EFT we use Eq.~\eqref{eq.25} at LO and Eq.~\eqref{eq.37} at NLO,
with their corresponding errors in Eqs. \eqref{eq.26} and \eqref{eq.38}.
We also compare the position of the poles determined numerically
in the toy model with the EFT predictions and their errors 
given in Sec. \ref{poles}.

\begin{figure}[t]
  \centering
  \includegraphics[scale=0.2]{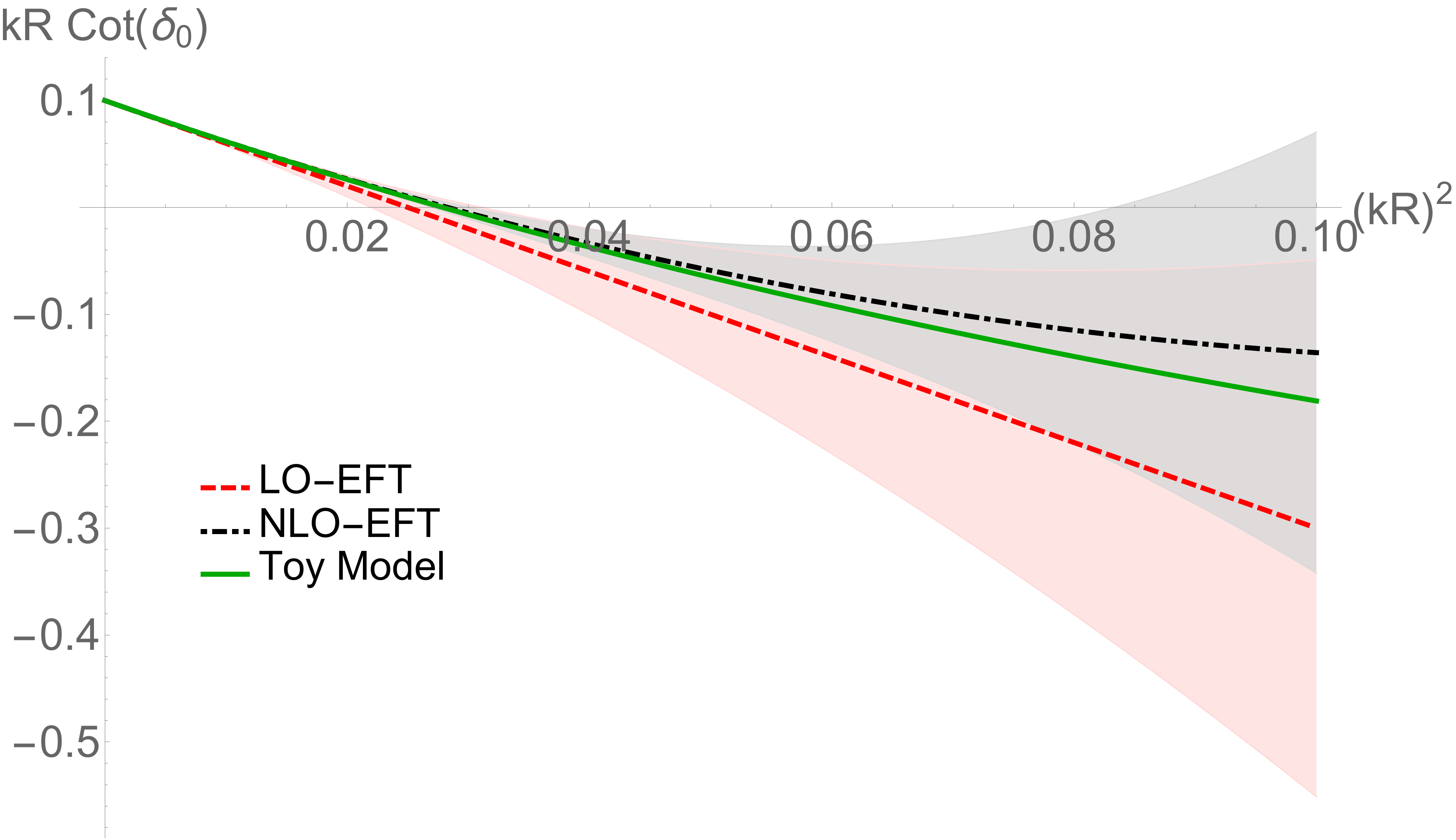}
  \includegraphics[scale=0.2]{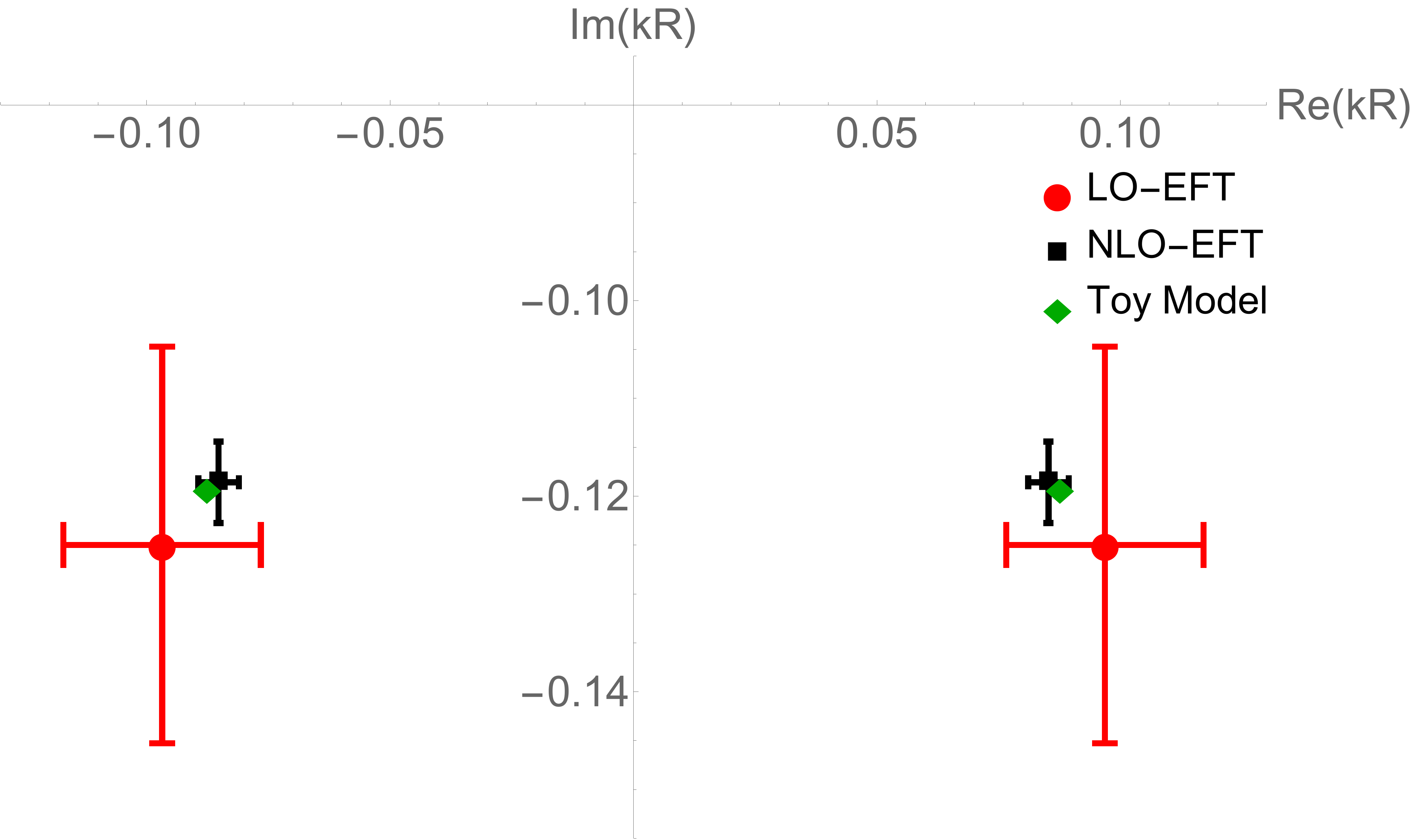}
\caption{Comparison between EFT and toy model for 
$kR \cot\delta_0$ as a function of $(kR)^2$ (left panel) 
and resonance pole positions on the complex $kR$ plane (right panel), when
$\alpha = 7.1555$ and $\beta^2 = 7.66264$.
On the left panel, the results for the EFT at LO and NLO
are indicated by, respectively, a (red) band around the (red) dashed line 
and a (gray) band around the (black) dot-dashed line,
while the toy model gives the (green) solid line.
On the right panel, the EFT at LO and NLO give, respectively,
the points with the larger (red) and smaller (black) error bars around the 
(green) point from the toy model.
}
\label{fig3}
\end{figure}

For illustration, we keep the effective range fixed at $r_{0}/R= -8$
and vary $a_{0}/R$ so as to reproduce the various cases discussed in the 
previous section.
We start with $a_{0}/R= -10 > 2r_{0}/R$.
The toy-model parameters for this choice are $\beta^2 = 7.66264$
and $\alpha = 7.15555$, which lead to $P_{0}=0.256439$.
On the left panel of Fig. \ref{fig3} we show the EFT phase shifts 
at LO and NLO compared to the toy-model values.
As expected, the LO EFT agrees with the toy model within an error that
increases with energy. The NLO perturbation improves
the agreement for the central value and the error is decreased compared to LO.
For this choice of parameters the EFT has two resonance poles,
just as the toy model.
The pole positions are shown on the right panel of Fig. \ref{fig3} 
and given in Table \ref{tbl2}.
Again the EFT error bars decrease with order and central values
approach the exact result.
This example demonstrates the power of EFT to approximate in a systematic and 
controlled way the $T$ matrix for resonant states.

\begin{table}[tb]
\begin{tabular}{c||c|c}
& $k_{1}R$ & $k_{2}R$ \\
\hline
\hline
LO EFT & $(0.10 \pm 0.02) - (0.13 \pm 0.02) \, i$ 
& $-(0.10 \pm 0.02) - (0.13 \pm 0.02) \, i$ 
\\
NLO EFT& $(0.085 \pm 0.004) - (0.119 \pm 0.004) \, i$ 
&$-(0.085 \pm 0.004) - (0.119 \pm 0.004) \, i$ 
\\
\hline
Toy model & $ 0.087583 - 0.119238 \, i$ 
&$- 0.087583 - 0.119238 \, i$ 
\end{tabular}
\caption{Position of resonance poles $k_{1,2}$ in units of $R^{-1}$, when
$\alpha = 7.1555$ and $\beta^2 = 7.66264$.
The EFT at LO and NLO is compared with the toy model.
}
\label{tbl2}
\end{table}

\begin{figure}[t]
  \centering
  \includegraphics[scale=0.2]{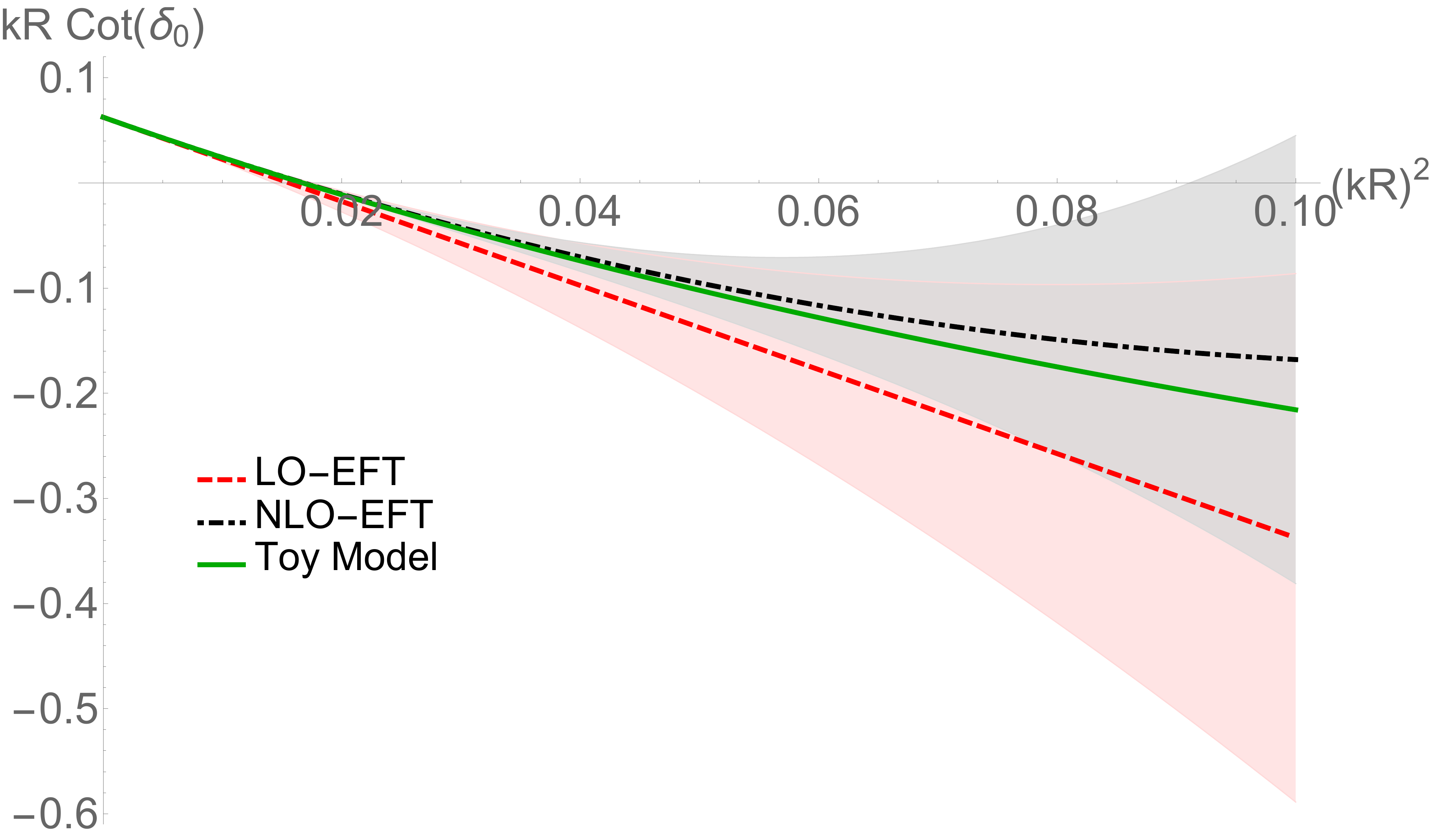}
  \includegraphics[scale=0.2]{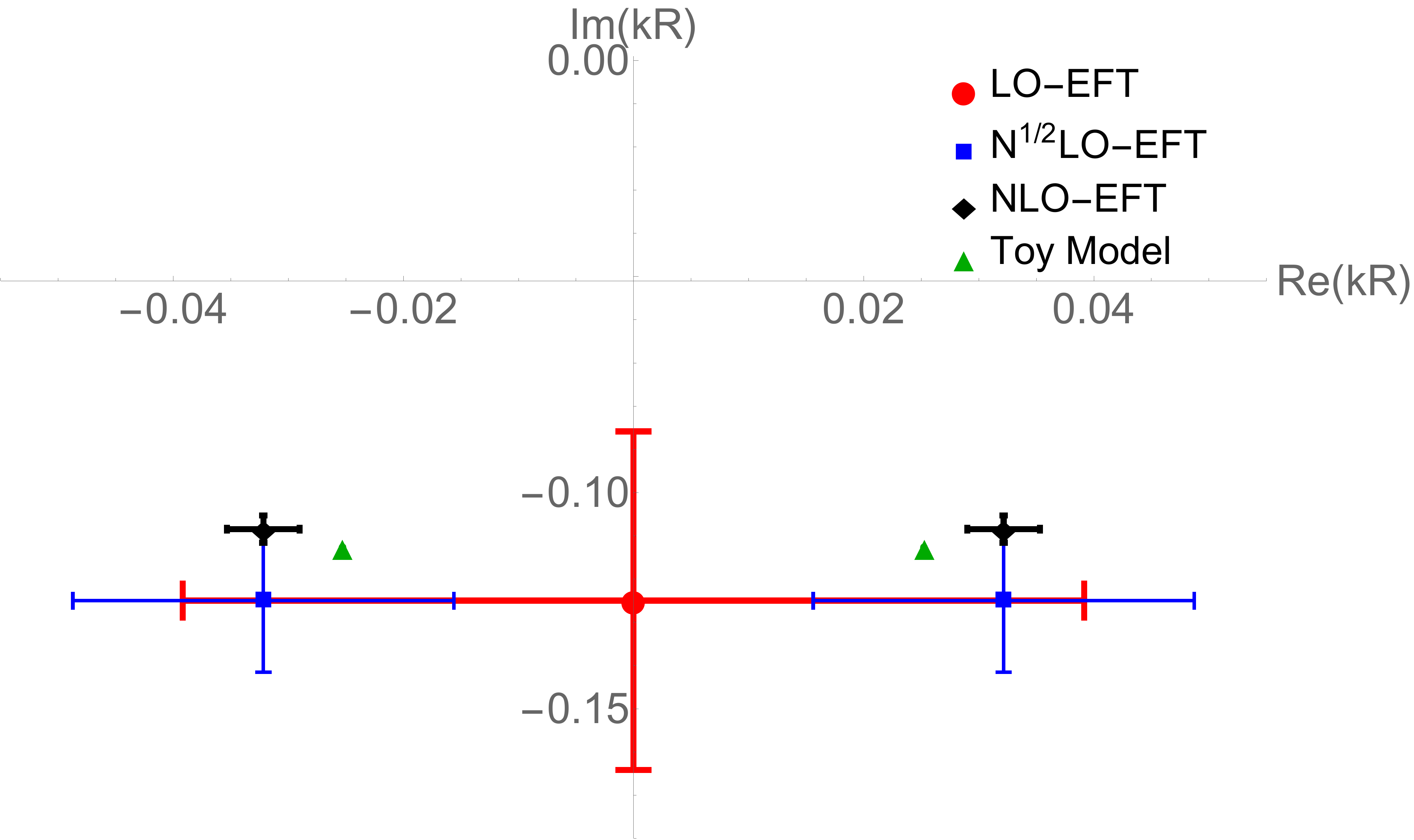}
\caption{Comparison between EFT and toy model for 
$kR \cot\delta_0$ as a function of $(kR)^2$ (left panel) 
and pole positions on the complex $kR$ plane (right panel), when
$\alpha = 7.4254$ and $\beta^2 = 7.7321$.
On the right panel, the EFT at N$^{1/2}$LO gives the (blue) point 
with intermediate-size error bar. 
Other notation as in Fig. \ref{fig3}.
}
\label{fig4}
\end{figure}

We now increase the attraction so that $a_{0}/R= -16 = 2r_{0}/R$,
when the toy-model parameters are $\beta^2 = 7.7321$
and $\alpha = 7.4254$. In this case $P_{0}=0.264856$.
The corresponding phase shifts are given on the left panel 
of Fig. \ref{fig4}. 
The larger magnitude of the scattering length means $k\cot\delta_0(k)$
is smaller at $k=0$, but the similar values of toy-model parameters 
lead to low-energy phase shifts that are not very different from
the previous case.
The pole positions
are given on the right panel of Fig. \ref{fig4} and in Table \ref{tbl3}.
The increased attraction brings the toy-model poles closer to the imaginary
axis.
The EFT has a double pole at LO with error bars that encompass the
two resonance poles. 
The half-power correction corrects for the horizontal splitting,
and the full NLO correction moves the poles slightly upwards.
The convergence pattern is clear, although the NLO error bars are
underestimated by a factor of about 2. 
Had we estimated them by simply multiplying
Eq. \eqref{eq.69} with $\sqrt{P_0}$, 
similarly to what we have done at N$^{1/2}$LO, 
the NLO error would be about four times larger.

\begin{table}[tb]
\begin{tabular}{c||c|c}
& $k_{1}R$ & $k_{2}R$ \\
\hline
\hline
LO EFT & $-(0.13 \pm 0.04)\, i$ & $-(0.13 \pm 0.04)\, i$
\\
N$^{1/2}$LO EFT& $(0.03 \pm 0.02) - (0.13 \pm 0.02) \, i$ 
&$-(0.03 \pm 0.02) - (0.13 \pm 0.02) \, i$ 
\\
NLO EFT& $(0.032 \pm 0.003) - (0.108 \pm 0.003)\, i$ 
&$-(0.032 \pm 0.003) - (0.108 \pm 0.003)\, i$
\\
\hline
Toy model & $0.025286  - 0.112685\, i$
& $- 0.025286  - 0.112685\, i$
\end{tabular}
\caption{Position of resonance poles $k_{1,2}$ in units of $R^{-1}$, when
$\alpha = 7.4254$ and $\beta^2 = 7.7321$.
The EFT at LO, N$^{1/2}$LO, and NLO is compared with the toy model.
}
\label{tbl3}
\end{table}

\begin{figure}[t]
  \centering
  \includegraphics[scale=0.2]{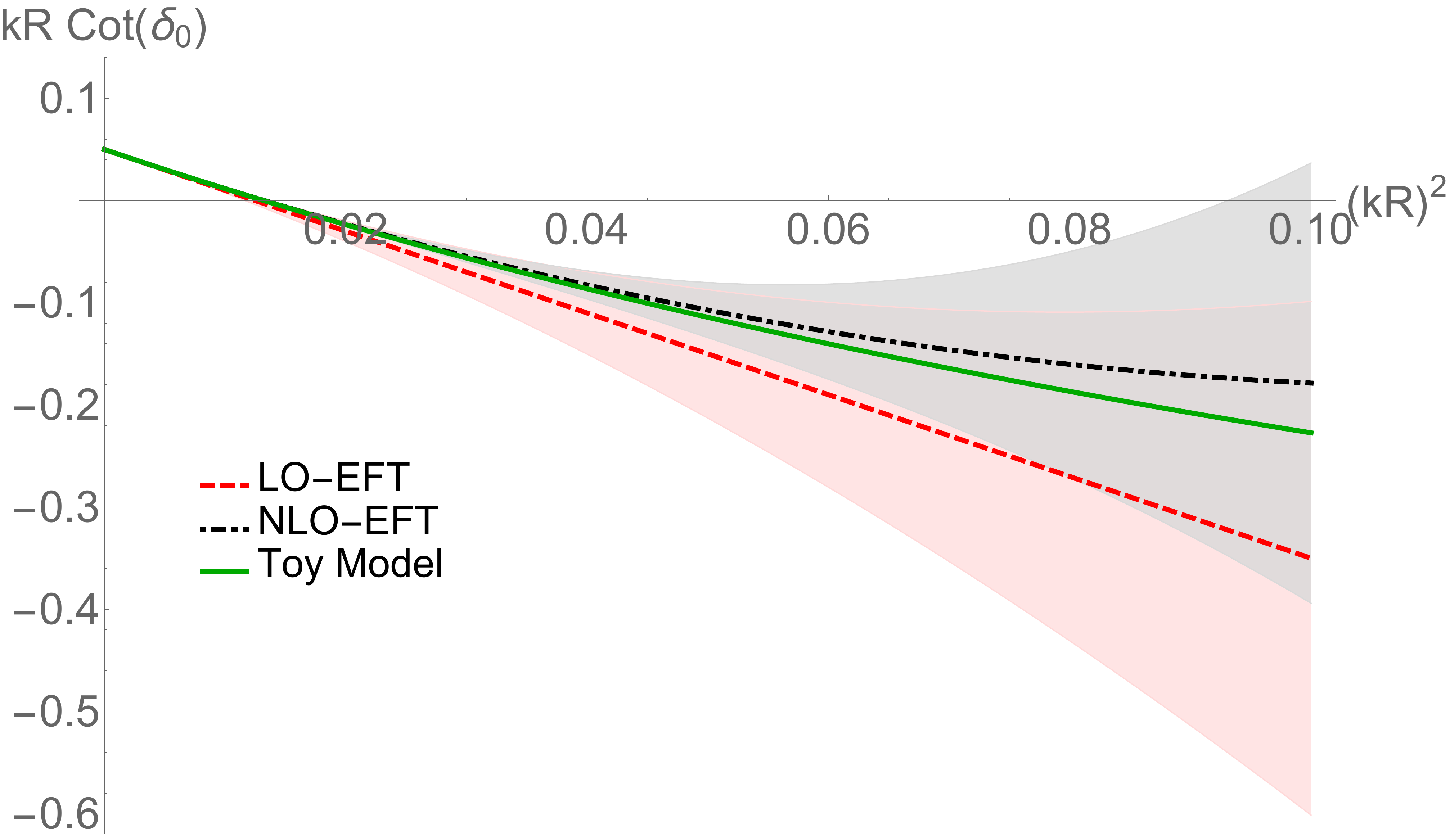}
  \includegraphics[scale=0.2]{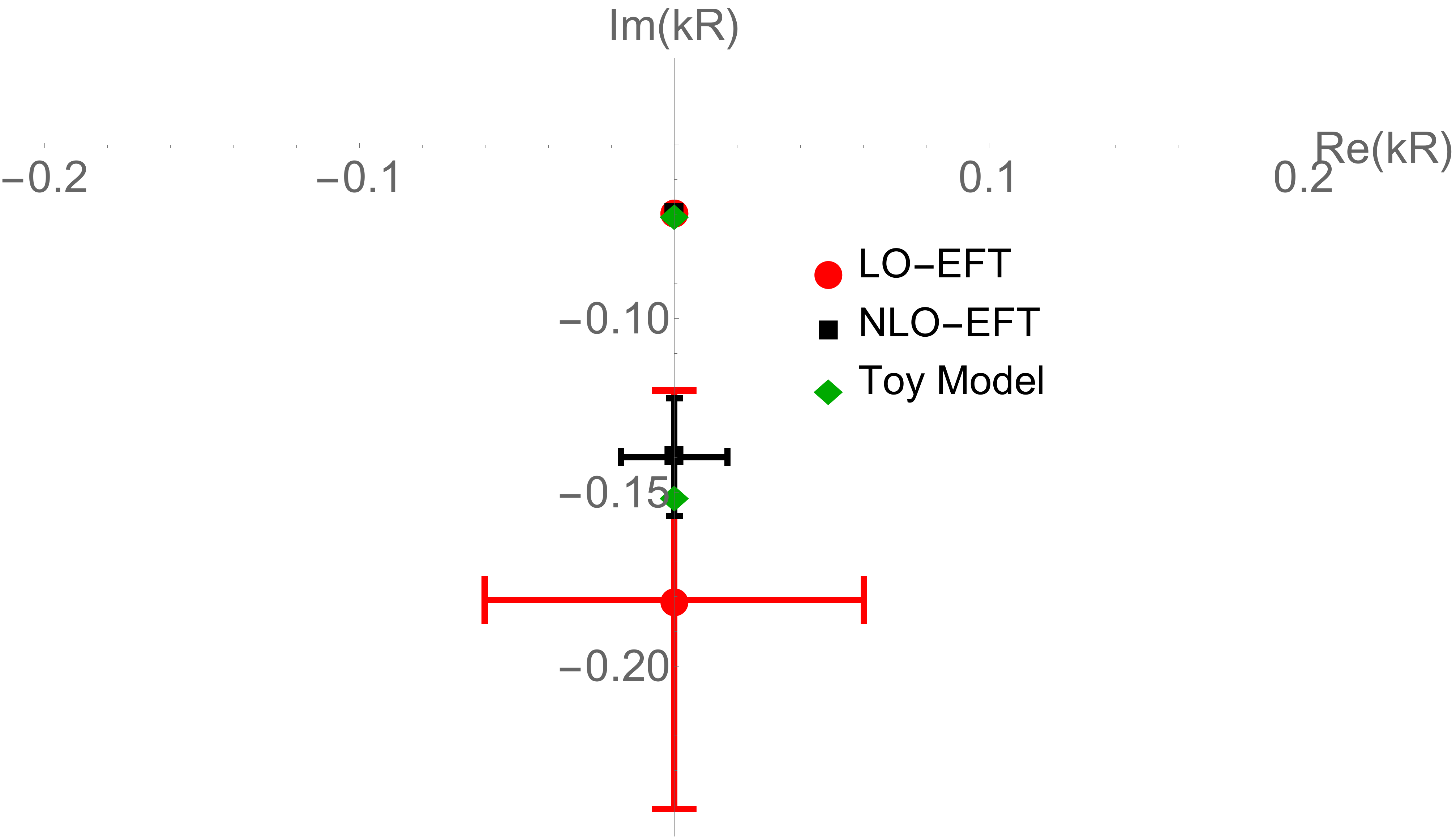}
\caption{Comparison between EFT and toy model for 
$kR \cot\delta_0$ as a function of $(kR)^2$ (left panel) 
and virtual-state pole positions on the complex $kR$ plane (right panel), when
$\alpha = 7.51962$ and $\beta^2 = 7.75532$.
Notation as in Fig. \ref{fig3}.
}
\label{fig5}
\end{figure}

Further increase of the attraction moves the two toy-model poles onto the 
negative imaginary axis, that is, it creates two virtual states.
Taking $a_{0}/R= -20 < 2r_{0}/R$,
the values for the potential parameters are $\beta^2=7.75532$ and 
$\alpha=7.51962$, with the shape parameter being 
$P_{0}=0.267795$.
The phase shifts  at low energies are seen on the left panel of 
Fig.~\ref{fig5}
to be, again, very similar to the previous cases.
The pole positions are nevertheless very different,
as shown on the right panel and in Table~\ref{tbl4}. 
As expected, the EFT describes the shallower pole very well, but has much
larger errors for the deeper state.

\begin{table}[tb]
\begin{tabular}{c||c|c}
& $k_{1}R$ & $k_{2}R$ \\
\hline
\hline
LO EFT & $-(0.069\pm 0.001)\, i $ & $-(0.18 \pm 0.06)\, i$ 
\\
NLO EFT& $-(0.0700\pm 0.0001)\, i$ & $-(0.14 \pm 0.02)\, i$ 
\\
\hline
Toy model & $-0.070047\, i $ & $-0.150999\, i $
\end{tabular}
\caption{Position of virtual poles $k_{1,2}$ in units of $R^{-1}$, when
$\alpha = 7.51962$ and $\beta^2 = 7.75532$.
The EFT at LO and NLO is compared with the toy model.
}
\label{tbl4}
\end{table}

\begin{figure}[t]
  \centering
  \includegraphics[scale=0.2]{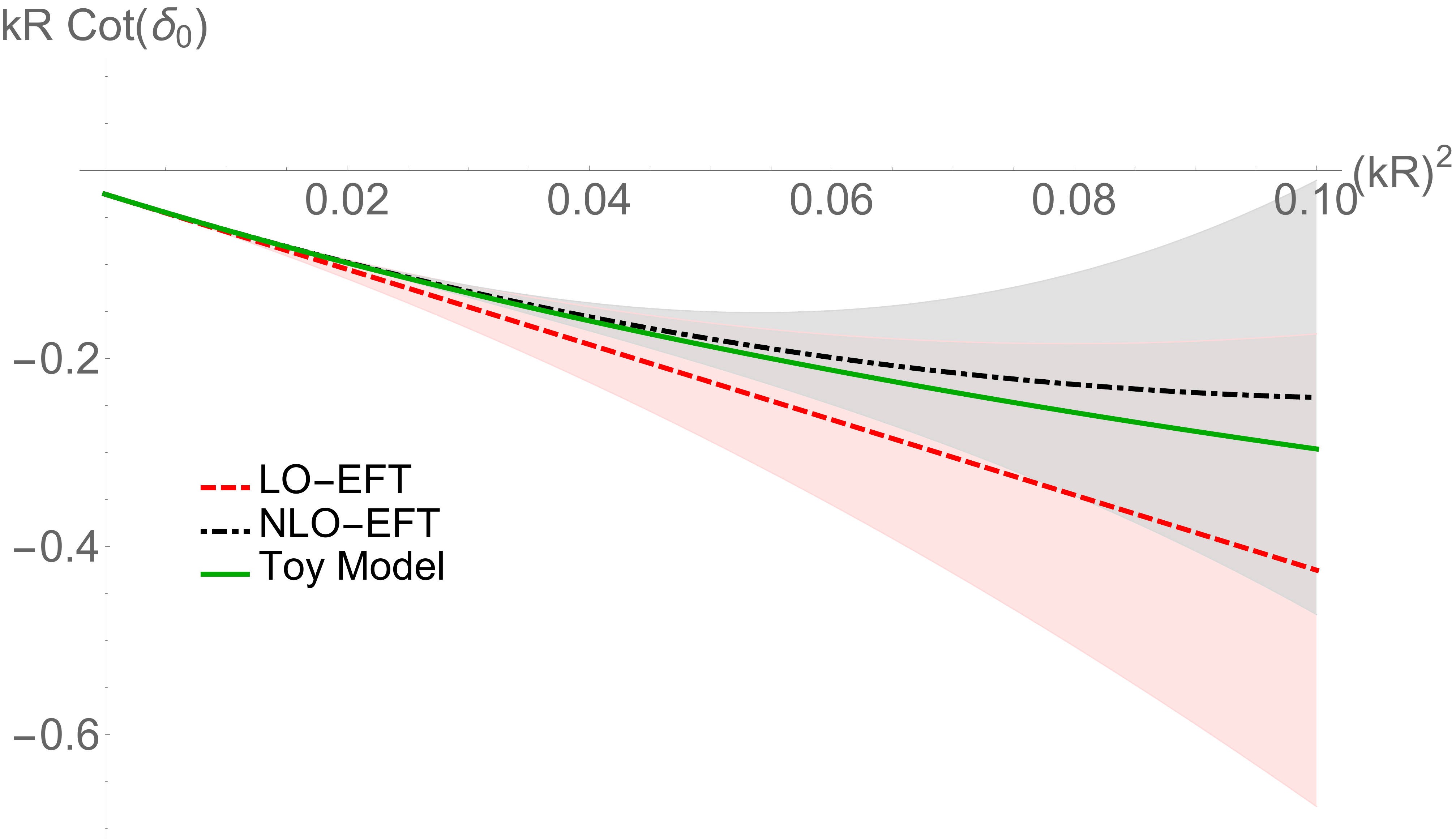}
  \includegraphics[scale=0.2]{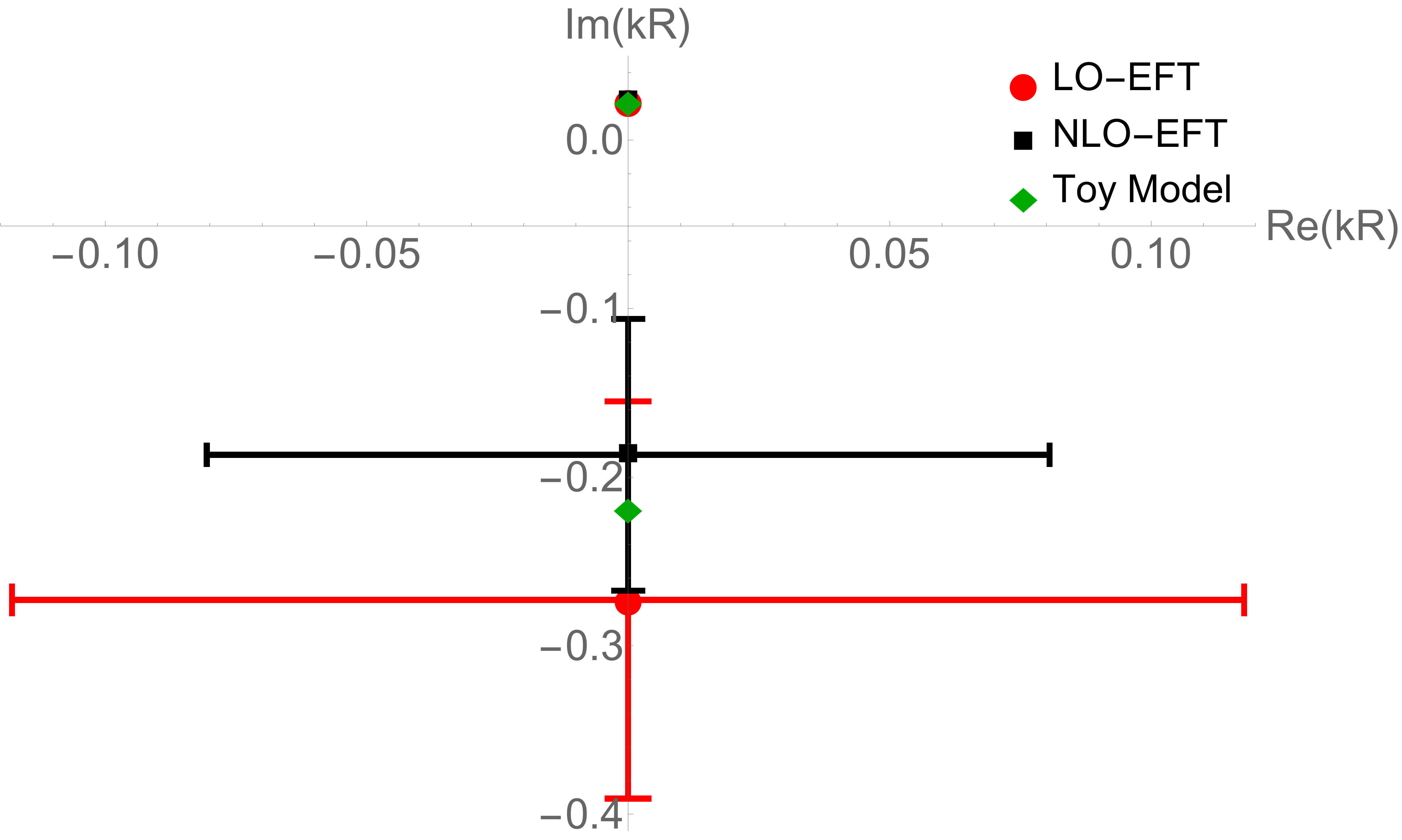}
\caption{Comparison between EFT and toy model for 
$kR \cot\delta_0$ as a function of $(kR)^2$ (left panel), 
and virtual- and bound-state 
pole positions on the complex $kR$ plane (right panel), when
$\alpha=8.13476$ and $\beta^2=7.89592$.
Notation as in Fig. \ref{fig3}.
}
\label{fig6}
\end{figure}

Finally, as an example of $a_{0}/R>0$, we take $a_{0}/R= 40$,
which translates into $\beta^2=7.89592$, $\alpha=8.13476$, and
$P_{0}=0.287009$. The phase shifts continues to resemble earlier cases, 
except that 
$a_{0} > 0$ changes the sign of the zero-energy value --- see the left panel
of Fig.~\ref{fig6}. The poles are shown on the right panel
of Fig.~\ref{fig6} and in Table \ref{tbl5}. 
Now there is a shallow bound state for which the EFT converges quickly.
The deeper, virtual state has even larger error bars than in
the previous example.

\begin{table}[tb]
\begin{tabular}{c||c|c}
& $k_{1}R$ & $k_{2}R$ \\
\hline
\hline
LO EFT & $(2.2902\pm 0.0006 )\times 10^{-2}\, i$ & $-(0.3 \pm 0.1)\, i$
\\
NLO EFT& $(2.289773 \pm 0.000003 )\times 10^{-2}\, i $ & $-(0.19 \pm 0.08)\, i$
\\
\hline
Toy model & $2.2897717\times 10^{-2}\, i$ & $-0.218497385\, i$
\end{tabular}
\caption{Position of bound and virtual poles $k_{1,2}$ in units of $R^{-1}$, when
$\alpha=8.13476$ and $\beta^2=7.89592$.
The EFT at LO and NLO is compared with the toy model.
}
\label{tbl5}
\end{table} 

These examples are sufficient to illustrate how in all two-pole configurations
the EFT reproduces the toy-model results with increased accuracy as
the order increases. 

\section{Conclusion}
\label{conc}

We have constructed an effective field theory that describes the scattering of 
two nonrelativistic particles when two shallow $S$-wave poles are present. 
Resonant nonrelativistic scattering has been considered before in EFT 
\cite{Bertulani:2002sz,Bedaque:2003wa,Higa:2008dn,Gelman:2009be, Alhakami:2017ntb}, 
but only with energy-dependent interactions. The characteristic feature of 
our formulation is the sole reliance on momentum-dependent interactions, 
which are easier to employ in more-particle systems. The effective Lagrangian 
for any low-energy two-body scattering process involving short-range forces 
can be written as an expansion in contact operators with an increasing number 
of spatial derivatives. The challenge, which we met above, is to order these 
interactions with a power counting appropriate to produce resonant poles in the 
scattering amplitude. 

It is well known that various observables --- like the scattering length and 
the effective range in the effective-range expansion --- reflect the presence 
of resonant states. This feature, which we have illustrated with a toy model 
against which we compare our EFT, motivates the power-counting scheme we use 
in this paper. We find that the low-energy scattering amplitude yields 
resonant poles only when the two leading operators in the derivative expansion 
are treated nonperturbatively. At higher orders, operators with an increasing 
number of derivatives contribute perturbatively. Our EFT successfully produces 
a controlled expansion about the resonant states and predicts the positions of 
poles with an error that can be systematically reduced by adding 
higher-dimensional operators. The same is true more generally for other 
situations involving two shallow $S$-wave poles, such as two virtual states, 
or one virtual state and one bound state.

As a {\it bona fide} EFT, ours obeys approximate renormalization-group 
invariance. It is remarkable that renormalization at leading order forces 
the effective range to be negative \cite{Phillips:1997xu,Beane:1997pk}. 
This is in agreement with Wigner's bound \cite{Wigner:1955zz} and allows 
no more than one pole in the upper half of the complex momentum plane. 
Thus, renormalization automatically incorporates the causality constraint 
that a resonance represents decaying, not growing, states 
\cite{Moller:1946,Hu:1948zz}. It also does not allow for a redundant pole on 
the positive imaginary axis \cite{Ma:1946,TerHaar:1946,Ma:1947zz} nor for 
a double bound-state pole, which is excluded by other arguments 
\cite{Moller:1946,Hu:1948zz,DemkovDrukarev1966}. In short, the resulting 
$S$ matrix obeys the conditions expected on general grounds \cite{Moller:1946}.
In contrast, the renormalization of the EFT with ``dimer'' auxiliary fields 
\cite{Bertulani:2002sz,Bedaque:2003wa,Higa:2008dn,Gelman:2009be, Alhakami:2017ntb} 
allows for the more general situation, believed to be unphysical, where two 
(or more) shallow poles appear in the upper half-plane.

The situation does not change at higher orders, since the corrections are 
perturbative. The need to remove the residual cutoff dependence gives clues 
about the orders corrections come at. We saw explicitly how the four-derivative
operator enters at next-to-leading order, gives rise to a known form for the 
$S$ matrix \cite{Hu:1948zz}, and improves on the leading-order description 
systematically. There is no obvious obstacle to continuing this process beyond 
next-to-leading order.

Power counting and renormalization here are significantly different than those 
\cite{vanKolck:1997ut,Kaplan:1998tg,Kaplan:1998we,vanKolck:1998bw} for a 
single shallow $S$-wave pole. The need to treat the two-derivative 
contact interaction perturbatively in the latter case has been known for 
a long time \cite{vanKolck:1997ut,Kaplan:1998tg,Kaplan:1998we,vanKolck:1998bw},
but now we understand this need from the fact that the situation described by 
a nonperturbative treatment of the two-derivative contact interaction is 
different in a physically meaningful way --- it corresponds to different 
regions of the parameter space of an underlying potential, or to a 
different type of potential altogether.  

The EFT developed in this paper is directly applicable only to resonant 
$S$-wave scattering of two 
spin-zero particles that interact via a short-range force. 
Although at the two-body level it is equivalent to a particular
ordering of the effective-range expansion, it is a
Hamiltonian framework that allows for the investigation of 
the effects of this two-body physics in processes 
involving more than two particles. This is analogous
to the single-pole theory, where for example 
the three-body system can be dealt with 
\cite{Bedaque:1997qi,Bedaque:1998mb,Bedaque:1998kg,Bedaque:1998km,Bedaque:1999ve}.
The next obvious step is to include spin and (in the nuclear case)
isospin quantum numbers, as well as higher waves.
Furthermore, we aim to study resonant scattering of electrically charged 
particles by including the long-range Coulomb interactions in our EFT. 
We anticipate that a nonperturbative treatment of the Coulomb interaction 
will be necessary to describe resonant scattering of alpha particles at 
low energies \cite{Higa:2008dn}. 
The EFT developed in this paper is a step in this direction.

\section*{Acknowledgments}

UvK is grateful to R. Higa for useful discussions.
This research was supported in part
by the U.S. Department of Energy, Office of Science,
Office of Nuclear Physics, under award number DE-FG02-04ER41338.

\appendix

\section{$T$ matrix from Feynman diagrams}
\label{AppxA}

\begin{figure}[t]
\begin{center}
\includegraphics[scale=.9]{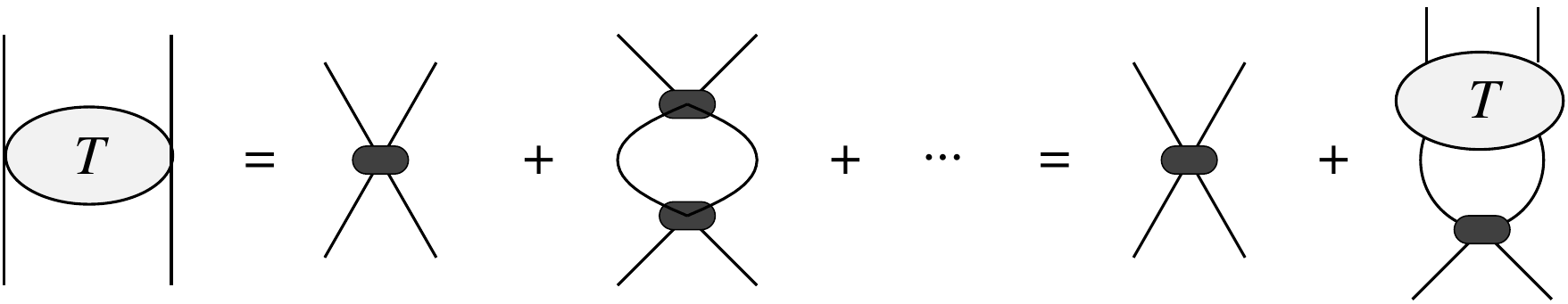}
\end{center}
\caption{The two-body $T$ matrix as a sum of Feynman diagrams.
Particle propagation \eqref{eqA1} is represented by a solid line,
while the dark oval stands for the potential \eqref{eq.3}.
}
\label{fig7}
\end{figure}

Here we obtain the $T$ matrix in field theory by summation of 
Feynman diagrams. The potential \eqref{eq.3} is represented by
four-legged vertices with an increasing number of powers of momenta.
Since antiparticles are integrated out, the two-body amplitude is just
a string of these vertices connected by two single-particle propagators,
\begin{equation}
S_{F}(q)= \frac{i}{q_0 - \vec{q}^{\,2}/2m + i \epsilon} +\ldots
\,,
\label{eqA1}
\end{equation}
where $q^0$ ($\vec{q}$) is the fourth (three-dimensional) component of the
particle's 4-vector and ``$\ldots$'' represent relativistic corrections.
We will neglect the latter here, but their inclusion poses no additional
conceptual problems \cite{vanKolck:1998bw}.
The sum of diagrams is shown in Fig. \ref{fig7}.

We can write the $S$-wave $T$ matrix in a compact way by
defining in ($\ket{p}$) and out ($\ket{p\rq{}}$) vectors and  
a vertex matrix ($\mathcal{C}$) through
\begin{equation}
\ket{p} \equiv 
\left(\begin{matrix} 1 \\ p^2 \\ p^4 \\ \vdots \end{matrix} \right) \,, 
\qquad 
\ket{p\rq{}} \equiv 
\left(\begin{matrix} 1 \\ p\rq{}^2 \\ p\rq{}^4 \\ \vdots \end{matrix} \right) 
\, ,
\qquad \mathcal{C} \equiv 
\frac{4\pi}{m}
\left(\begin{matrix} 
C_{0}   & C_{2}/2  & C_{4}/4  & \cdots\\ 
C_{2}/2 & C_{4}/2  & 3C_{6}/8 & \cdots\\ 
C_{4}/4 & 3C_{6}/8  & 3C_{8}/8 & \cdots\\
\vdots & \vdots & \vdots & \ddots
\end{matrix}\right) \, .
\label{eqA2}
\end{equation}
(See also Ref. \cite{Beck:2019abp}.)
The tree diagram in Fig. \ref{fig7} is then simply
\begin{equation}
T_{0;0}(p\rq{},p) = V(p\rq{},p) 
= \bra{p\rq{}}\,\mathcal{C}\,\ket{p} \,.
\label{eqA3}
\end{equation}

The loop diagrams involve a 4-momentum integration.
Since the vertices depend only on the 3-momenta,
we can evaluate the $q^{0}$ integrals in the center-of-mass frame,
\begin{equation}
\int\!\!\frac{d^4 q}{(2 \pi)^4}\, \vec{q}^{\,2n} \,
\frac{i}{q^{0} - \vec{q}^{\,2}/2m + i \epsilon}
\,\frac{i}{k^2/2m - q^{0} - \vec{q}^{\,2}/2m + i \epsilon} 
= -i \frac{m}{4\pi} I^{+}_{2n}
\, ,
\label{eqA4}
\end{equation}
where $I^{+}_{2n}$ is defined in Eq. \eqref{eq.12}.
If we define a matrix of integrals,
\begin{equation}
\mathcal{I} \equiv - m \int\!\!\frac{d^3q}{(2 \pi)^3} 
\frac{\ket{q}\bra{q}}{k^2-q^2+i\epsilon}
= -\frac{m}{4\pi}\left(\begin{matrix} 
I_{0}^{+} & I_{2}^{+} & I_{4}^{+} & \cdots\\
I_{2}^{+} & I_{4}^{+} & I_{6}^{+} & \cdots\\
I_{4}^{+} & I_{6}^{+} & I_{8}^{+} & \cdots\\
\vdots & \vdots & \vdots & \ddots
\end{matrix}\right) \, ,
\label{eqA5}
\end{equation}
the one-loop diagram takes the form
\begin{equation}
T_{0;1}(p\rq{},p) = - m \int\!\!\frac{d^3q}{(2 \pi)^3} 
\frac{V(p\rq{},q)\,V(q,p)}{k^2-q^2+i\epsilon}
=\bra{p\rq{}}\,\mathcal{C}\,\mathcal{I}\,\mathcal{C}\,\ket{p}
\,.
\label{eqA6}
\end{equation}
With a regulator on nucleon momenta, the multiple-loop diagrams
separate and the sum of diagrams is
\begin{equation}
T_{0}(p\rq{},p) =\sum_{i=0}^{\infty} T_{0;i}(p\rq{},p)
= \bra{p\rq{}}
\left(\mathcal{C} + \mathcal{C}\,\mathcal{I}\,\mathcal{C} 
+ \mathcal{C}\,\mathcal{I}\,\mathcal{C}\,\mathcal{I}\,\mathcal{C} + \ldots
\right)\ket{p}
= \bra{p\rq{}}\,\mathcal{C}\left(1 - \mathcal{I}\,\mathcal{C}\right)^{-1}\ket{p}
\equiv \bra{p\rq{}}\,\mathcal{T}_{0}\,\ket{p}
\,.
\label{eqA7}
\end{equation}

Since predictive power requires a finite number of parameters at each order,
we need to truncate the amplitude \eqref{eqA7} 
at different orders in order to renormalize it.
The LO $T$ matrix results from taking 
$C_0 = C_{0}^{(0)}$, $C_2 = C_{2}^{(0)}$, and $C_{n\ge 4}=0$ 
in the matrix $\mathcal{C}$.
We arrive at 
\begin{equation}
\frac{4\pi}{mT^{(0)}_{0}(p\rq{},p)} =  
\frac{(4C_{0}^{(0)}-C_{2}^{(0)2}I_{4})\, I_{0} + (2+C_{2}^{(0)} I_{2})^{2}}
      {4C_{0}^{(0)}-C_{2}^{(0)2}I_{4}+C_{2}^{(0)}(2+C_{2}^{(0)}I_{2})(p\rq{}^{2}+p^{2}) 
       -C_{2}^{(0)2}I_{0} \, p\rq{}^{2}p^{2}}
\,,
\label{eqA8}
\end{equation}
which gives Eq.~\eqref{eq.16} on-shell, that is, when $p^2 = p\rq{}^2 = k^2$.
To obtain the NLO $T$ matrix,
we take instead $C_0 = C_{0}^{(0)} + C_{0}^{(1)}$, $C_2 = C_{2}^{(0)} + C_{2}^{(1)}$,
$C_{4} = C_{4}^{(1)}$, and $C_{n\ge 6}=0$,
expanding in the subleading pieces. 
Retaining only terms linear in $C_{0,2,4}^{(1)}$ results on-shell 
in Eq. \eqref{eq.29}.
The procedure can be continued in a straightforward way
at higher orders.

Note that the same results can be obtained from Feynman diagrams
in a form that is somewhat closer to the procedure of the main text.
We rewrite the sum of diagrams \eqref{eqA7} as a Lippmann-Schwinger equation,
\begin{equation}
\bra{p\rq{}}\,\mathcal{T}_{0}\,\ket{p}
=\bra{p\rq{}}\,\mathcal{C}\,\ket{p} 
+ \bra{p\rq{}}\,\mathcal{C}\,\mathcal{I}\,\mathcal{T}_{0}\,\ket{p}
\,.
\label{eqA9}
\end{equation}
This form is also shown in Fig. \ref{fig7}.
The Lippmann-Schwinger equation can be solved at LO \cite{Phillips:1997xu}
with an {\it ansatz} motivated by the
momentum structure of Eq. \eqref{eqA6},
\begin{equation}
\mathcal{T}_{0}^{(0)} =
\left(\begin{matrix} 
\tau_0^{(0)} & \tau_2^{(0)} & 0 & \cdots\\
\tau_2^{(0)} & \tau_4^{(0)} & 0 & \cdots\\
0 & 0 & 0 & \cdots\\
\vdots & \vdots & \vdots & \ddots
\end{matrix}
\right)
\, .
\label{eqA10}
\end{equation}
Inserting this form on both sides of Eq. \eqref{eqA9}
and matching powers of momenta, one finds three algebraic equations
for $\tau_{0,2,4}^{(0)}$.
Solving these equations we again obtain Eq. \eqref{eqA8}.
At NLO, we expand both ${\cal C}$ and ${\cal T}_0$ in Eq. \eqref{eqA9}
to linear order.
The NLO correction ${\cal T}_0^{(1)}$ appears both directly on the 
left-hand side and inside an integral on the right-hand side.
To solve the resulting equation we make an {\it ansatz} 
analogous to \eqref{eqA10} but now including the sixth power of momenta.

\section{Renormalization procedure} 
\label{AppxB}

In this appendix we give some of the details of our renormalization
procedure, at both LO and NLO.
In either case we expand the amplitude calculated within the EFT 
in a power series in $k/ \Lambda \ll 1$ as
\begin{equation}
\frac{4\pi}{mT_{0}(k)} = ik + \sum_{n = 0}^\infty g_{2n}k^{2n} \, ,
\label{eqB1}
\end{equation}
where the coefficients 
\begin{equation}
g_{2n}=g_{2n}^{(0)}+g_{2n}^{(1)}+\ldots
\label{eqB2}
\end{equation}
depend on the bare LECs $C_{2n}(\Lambda)$
and the cutoff $\Lambda$. Since $g_{2n}$ encodes short-range physics,
only integer powers of the energy appear in the expansion \eqref{eqB1}.
The only non-analytic behavior is represented by the unitarity term
$ik$, which stems from the Schr\"odinger propagation.
The $C_{2n}(\Lambda)$ are fixed by matching Eq. \eqref{eqB1}
with the ERE in Eq.~\eqref{eq.1}.

At LO, the amplitude is given by Eq.~\eqref{eq.16}. 
The first four non-zero $g_{2n}^{(0)}$ are:
\bea
g_0^{(0)} & = & L_{1} 
-\frac{1}{C_{2}^{(0)2}L_{5}-4C_{0}^{(0)}} 
\left(C_{2}^{(0)}L_{3}+2\right)^2
\, ,
\label{eqB3}
\\
g_2^{(0)} & = & L_{-1} 
- C_{2}^{(0)}\frac{C_{2}^{(0)}L_{3}+4}{(C_{2}^{(0)2}L_{5}-4C_{0}^{(0)})^2} 
\left(C_{2}^{(0)}L_{3}+2\right)^2
\, ,
\label{eqB4} 
\\
g_4^{(0)} & = & L_{-3} 
- C_{2}^{(0)2}\frac{(C_{2}^{(0)}L_{3}+4)^2}{(C_{2}^{(0)2}L_{5}-4C_{0}^{(0)})^3} 
\left(C_{2}^{(0)}L_{3}+2\right)^2
\, ,
\label{eqB5}
\\
g_6^{(0)} & = & L_{-5} 
-C_{2}^{(0)3} \frac{(C_{2}^{(0)}L_{3}+4)^3}
{(C_{2}^{(0)2}L_{5}-4C_{0}^{(0)})^4} 
\left(C_{2}^{(0)}L_{3}+2\right)^2
\, .
\label{eqB6}
\eea
The expressions for $g_{0,2}^{(0)}$ agree with those in Refs. 
\cite{Phillips:1997xu,Beane:1997pk}.
We demand that $g_{0,2}^{(0)}$ reproduce given values of the scattering
length $a_0$ and effective range $r_0$, 
\bea
g_0^{(0)} & = & \frac{1}{a_0}\, ,
\label{eqB7}
\\
g_2^{(0)} & = & -\frac{r_0}{2} \, .
\label{eqB8} 
\eea
This is a set of two equations from which 
the two running values of $C_{0}^{(0)}$ and $C_{2}^{(0)}$ 
can be obtained in terms of the $L_{n}$ 
and the observables $a_0$ and $r_{0}$ as
\bea
C_{0}^{(0)} &=& 
- \frac{L_5}{L_{3}^2}
\left[
1\mp 
\frac{\sqrt{2}\left(a_{0} L_{1}-1\right)}
{\sqrt{-a_{0}^2(r_{0}+2L_{-1})L_{3}+2(a_{0}L_{1}-1)^2}}
\right]^2
+\frac{2a_{0}(a_{0}L_{1}-1)}{a_{0}^2(r_{0}+2L_{-1})L_{3}-2(a_{0}L_{1}-1)^2} 
\, ,
\label{eqB9}\\ 
\notag \\
C_{2}^{(0)} &=&-\frac{2}{L_{3}}\left[
1\mp \frac{\sqrt{2}\left(a_{0}L_{1} - 1\right)}
{\sqrt{-a_{0}^2(r_{0}+2L_{-1})L_{3}+2(a_{0}L_{1}-1)^2}} 
\right] \, .
\label{eqB10}
\eea
Equations \eqref{eq.19} and \eqref{eq.20} follow upon 
expanding the expressions above for $a_0\Lambda\gg 1$ and $r_0\Lambda\gg 1$.
In addition, $g_4^{(0)}$ gives the residual dependence shown in 
Eq. \eqref{eq.23}.

At NLO, the amplitude is given by Eq. \eqref{eq.29}.
Once we expand in $k/\Lambda$ the shifts
in the first four $g_{2n}$ are given by
\bea
g_{0}^{(1)} & = & 
-\frac{4}{(C_{2}^{(0)2} L_{5}-4C_{0}^{(0)})^2} 
\left\{
C_{0}^{(1)} \left(C_{2}^{(0)} L_{3}+2\right)^2 
-C_{2}^{(1)} \left(C_{2}^{(0)} L_{3}+2\right)
\left(C_{2}^{(0)} L_{5}+2C_{0}^{(0)} L_{3}\right) 
\right. 
\notag \\
& & \left. 
+\frac{C_{4}^{(1)}}{4} 
\left[C_{2}^{(0)2}\left(C_{2}^{(0)} L_{3}+4\right)
\left(L_{5}^2-L_{3} L_{7}\right) 
- 4 C_{2}^{(0)} L_{7} 
+ 4 C_{0}^{(0)} L_{5} \left(C_{2}^{(0)} L_{3}-2\right) 
+ 8 C_{0}^{(0)2} L_{3}^2 
\right] 
\right\} \, ,
\label{eqB11} \\
\notag \\
g_{2}^{(1)} & = & 
\frac{8(C_{2}^{(0)} L_{3}+2)}{(C_{2}^{(0)2} L_{5}-4C_{0}^{(0)})^3}
\Biggl\{
C_{0}^{(1)}C_{2}^{(0)}\left(C_{2}^{(0)}L_{3}+2\right)\left(C_{2}^{(0)}L_{3}+4\right)
\notag \\
&& 
+ C_{2}^{(1)} 
\left[
C_{2}^{(0)}\left(C_{2}^{(0)}L_{3}+4\right)\left(C_{2}^{(0)}L_{5}+2C_{0}^{(0)}L_{3}\right)
- C_{2}^{(0)2}L_{5}+4C_{0}^{(0)}
\right]
\notag \\
&&
+ \frac{C_{4}^{(1)}}{4} 
\biggl[
C_{2}^{(0)3}\left(C_{2}^{(0)}L_{3}+4\right)\left(L_{5}^2 - L_{3} L_{7}\right)
+ 2C_{2}^{(0)2} L_{7} \left(C_{2}^{(0)}L_{3}+4\right) 
- 2C_{0}^{(0)} C_{2}^{(0)2} L_{3} L_{5}
\notag \\
&&
\qquad \qquad  
- 8C_{0}^{(0)2} L_{3} \left(C_{2}^{(0)} L_{3}+3\right) 
\biggr] 
\Biggr\} \, , 
\label{eqB12} \\
\notag \\
g_{4}^{(1)} & = & 
-\frac{4(C_{2}^{(0)}L_{3}+2)(C_{2}^{(0)}L_{3}+4)}{(C_{2}^{(0)2}L_{5}-4C_{0}^{(0)})^4}
\Biggl\{
3C_{0}^{(1)}C_{2}^{(0)2}\left(C_{2}^{(0)}L_{3}+2\right)\left(C_{2}^{(0)}L_{3}+4\right)
\notag \\
& & 
-C_{2}^{(1)}C_{2}^{(0)} 
\left[3C_{2}^{(0)} \left(C_{2}^{(0)}L_{3}+4\right)
\left(C_{2}^{(0)}L_{5}+2C_{0}^{(0)}L_{3}\right)
-4\left(C_{2}^{(0)2}L_{5}-4C_{0}^{(0)}\right)
\right]
\notag \\
& & 
+\frac{C_{4}^{(1)}}{4} 
\biggl[ 
3C_{2}^{(0)4} \left(C_{2}^{(0)} L_{3}+6\right)\left(L_{5}^2 - L_{3} L_{7}\right) 
-8C_{2}^{(0)3} L_{7} 
+4C_{2}^{(0)3}L_{5} \left(C_{0}^{(0)} L_{3}-2C_{2}^{(0)}L_{5}\right)  
\notag \\ 
&&\qquad\qquad 
+32 C_{0}^{(0)2} \left(C_{2}^{(0)} L_{3}+1\right) 
+24 C_{0}^{(0)2} C_{2}^{(0)} L_{3} \left(C_{2}^{(0)} L_{3}+2\right) 
\biggr] 
\Biggr\} \, ,
\label{eqB13} \\
\notag \\
g_{6}^{(1)} & = &
-\frac{8(C_{2}^{(0)}L_{3}+2)(C_{2}^{(0)}L_{3}+4)^2}{(C_{2}^{(0)2}L_{5}-4C_{0}^{(0)})^5}
C_{2}^{(0)} 
\Biggl\{ 
4C_{0}^{(1)}C_{2}^{(0)2}\left(C_{2}^{(0)}L_{3}+2\right)\left(C_{2}^{(0)}L_{3}+4\right)
\notag \\
& & 
-C_{2}^{(1)} C_{2}^{(0)} 
\left[
2C_{2}^{(0)}\left(C_{2}^{(0)}L_{3}+4\right)\left(C_{2}^{(0)}L_{5}+2C_{2}^{(0)}L_{3}\right)
-3\left(C_{2}^{(0)2}L_{5}-4C_{0}^{(0)}\right)
\right]
\notag \\
& & 
+\frac{C_{4}^{(1)}}{2} 
\biggl[ 
C_{2}^{(0)4} \left(C_{2}^{(0)}L_{3}+6\right)\left(L_{5}^2 - L_{3} L_{7}\right) 
+ 8 C_{2}^{(0)3} L_{7} 
- C_{2}^{(0)3}L_{5} \left(C_{0}^{(0)} L_{3}-3C_{2}^{(0)}L_{5}\right)
\notag \\ 
&&\qquad\qquad
+ 16 C_{0}^{(0)2} \left(C_{2}^{(0)} L_{3}+1\right) 
+ 4 C_{0}^{(0)2}C_{2}^{(0)}L_{3}\left(2C_{2}^{(0)}L_{3}+3\right)
\biggr]
\Biggr\}. 
\label{eqB14}
\eea
Now we demand that the shape parameter $P_0$ be reproduced,
without changes in the scattering length and effective ranges; that is,
we impose
\bea
g_{0}^{(1)} & = & 0 \, , 
\label{eqB15}
\\
g_{2}^{(1)} & = & 0 \, , 
\label{eqB16}
\\
g_{4}^{(1)} & = & P_0 \left(\frac{r_0}{2}\right)^3- g_{4}^{(0)} \, .
\label{eqB17}
\eea
Solving for the three unknowns $C_{0,2,4}^{(1)}$ which appear linearly,
\bea
C_{0}^{(1)} &=& -\left(\frac{P_0r_0^3}{8}-g_{4}^{(0)}\right)
\frac{(C_{2}^{(0)2}L_{5}-4C_{0}^{(0)})^2}{16(C_{2}^{(0)} L_{3}+2)^{2}(C_{2}^{(0)} L_{3}+4)} 
\biggl[
C_{2}^{(0)2} \left(C_{2}^{(0)} L_{3}+4\right)\left(L_{5} ^{2}+L_{3} L_{7}\right)
+ 4C_{2}^{(0)}L_{7}
\notag \\
&& \qquad  
+ 6 C_{0}^{(0)} C_{2}^{(0)} L_{3} L_{5} \left(C_{2}^{(0)}L_{3}+4\right)
+ 4 C_{0}^{(0)2} L_{3}^{2} \left(C_{2}^{(0)} L_{3}+4\right) 
+ 8 C_{0}^{(0)}L_{5} 
\biggr] 
\, ,
\label{eqB18}
\\ 
\notag \\
C_{2}^{(1)} &=& -\left(\frac{P_0r_0^3}{8}-g_{4}^{(0)}\right)
\frac{(C_{2}^{(0)2} L_{5}-4C_{0}^{(0)})^2}{8(C_{2}^{(0)} L_{3}+2)(C_{2}^{(0)} L_{3}+4)} 
\left[ 
C_{2}^{(0)} L_{5} \left(C_{2}^{(0)} L_{3}+4\right)
+C_{0}^{(0)} L_{3} \left(C_{2}^{(0)} L_{3}+6\right)
\right]
\, ,
\label{eqB19}
\\ 
\notag \\
C^{(1)}_{4} &=& -\left(\frac{P_0r_0^3}{8}-g_{4}^{(0)}\right)
\frac{(C_{2}^{(0)2} L_{5}-4C_{0}^{(0)})^2}{8(C_{2}^{(0)} L_{3}+4)}
\, ,
\label{eqB20}
\eea
where $g_{0}^{(0)}$, $C_{0}^{(0)}$ and $C_{2}^{(0)}$ are given in 
Eqs. \eqref{eqB5}, \eqref{eqB9} and \eqref{eqB10}. 
Expanding these expressions for $a_0\Lambda\gg 1$ and $r_0\Lambda\gg 1$
we obtain Eqs.~\eqref{eq.33}, \eqref{eq.34}, and \eqref{eq.35}.
The residual cutoff dependence in Eq. \eqref{eq.36} results from 
$g_{6}^{(0)}+g_{6}^{(1)}$.


\begin{thebibliography}{99}

\bibitem{Moller:1946}
C. M\o ller,
Kgl. Danske Vid. Selsk. Mat.-Fys. Medd. {\bf 22} (1946) 19.

\bibitem{TaylorScattering}
J.R Taylor, {\it Scattering Theory: The Quantum Theory of Nonrelativistic
Collisions}, Wiley, New York (1972).

\bibitem{Kok:1980dh}
  L.P.~Kok,
  Phys. Rev. Lett. {\bf 45} (1980) 427.

\bibitem{AFZAL:1969zz}
  S.A.~Afzal, A.A.Z.~Ahmad, and S.~Ali,
  Rev. Mod. Phys. {\bf 41} (1969) 247.

\bibitem{Weinberg:1978kz}
  S.~Weinberg,
  Physica A {\bf 96} (1979) 327.

\bibitem{Weinberg:1979pi}
  S.~Weinberg,
  Rev. Mod. Phys. {\bf 52} (1980) 515
  [Science {\bf 210} (1980) 1212].

\bibitem{Bedaque:2002mn} 
  P.F.~Bedaque and U.~van Kolck,
  Ann. Rev. Nucl. Part. Sci. {\bf 52} (2002) 339.

\bibitem{Hammer:2019poc}
H.-W.~Hammer, S.~K\"onig, and U.~van Kolck,
Rev. Mod. Phys. \textbf{92} (2020) 025004.
 
\bibitem{vanKolck:1997ut}
  U.~van Kolck,
  Lect. Notes Phys. {\bf 513} (1998) 62.

\bibitem{Kaplan:1998tg}
  D.B.~Kaplan, M.J.~Savage, and M.B.~Wise,
  Phys. Lett. B {\bf 424} (1998) 390.

\bibitem{Kaplan:1998we}
  D.B.~Kaplan, M.J.~Savage, and M.B.~Wise,
  Nucl. Phys. B {\bf 534} (1998) 329.

\bibitem{vanKolck:1998bw}
  U.~van Kolck,
  Nucl. Phys. A {\bf 645} (1999) 273.

\bibitem{Kaplan:1996nv}
  D.B.~Kaplan,
  Nucl. Phys. B {\bf 494} (1997) 471.

\bibitem{Bertulani:2002sz}
  C.A.~Bertulani, H.-W.~Hammer, and U.~van Kolck,
  Nucl. Phys. A {\bf 712} (2002) 37.

\bibitem{Bedaque:2003wa}
  P.F.~Bedaque, H.-W.~Hammer, and U.~van Kolck,
  Phys. Lett. B {\bf 569} (2003) 159.

\bibitem{Higa:2008dn}
  R.~Higa, H.-W.~Hammer, and U.~van Kolck,
  Nucl. Phys. A {\bf 809} (2008) 171.

\bibitem{Gelman:2009be}
  B.A.~Gelman,
  Phys. Rev. C {\bf 80} (2009) 034005.

\bibitem{Alhakami:2017ntb}
  M.H.~Alhakami,
  Phys. Rev. D {\bf 96} (2017) 
  056019.

\bibitem{Rotureau:2012yu}
  J.~Rotureau and U.~van Kolck,
  Few-Body Syst. {\bf 54} (2013) 725.

\bibitem{Ji:2014wta}
  C.~Ji, C.~Elster, and D.R.~Phillips,
  Phys. Rev. C {\bf 90} (2014) 
  044004.

\bibitem{Ryberg:2017tpv}
  E.~Ryberg, C.~Forss\'en, and L.~Platter,
  Few-Body Syst. {\bf 58} (2017) 
  143.

\bibitem{Bethe:1949yr}
  H.A.~Bethe,
  Phys. Rev. {\bf 76} (1949) 38.

\bibitem{vanKolck:2020plz}
  U.~van Kolck,
  Eur. Phys. J. A \textbf{56} (2020) 
  97.

\bibitem{Phillips:1997xu}
  D.R.~Phillips, S.R.~Beane, and T.D.~Cohen,
  Annals Phys. {\bf 263} (1998) 255.

\bibitem{Beane:1997pk}
  S.R.~Beane, T.D.~Cohen, and D.R.~Phillips,
  Nucl. Phys. A {\bf 632} (1998) 445.

\bibitem{Wigner:1955zz}
  E.P.~Wigner,
  Phys. Rev. {\bf 98} (1955) 145.

\bibitem{Beck:2019abp}
S.~Beck, B.~Bazak, and N.~Barnea,
Phys. Lett. B \textbf{806} (2020) 135485.

\bibitem{Cohen:2004kf}
  T.D.~Cohen, B.A.~Gelman and U.~van Kolck,
  Phys. Lett. B {\bf 588} (2004) 57.

\bibitem{Georgi:1990um}
  H.~Georgi,
  Phys. Lett. B {\bf 240} (1990) 447.

\bibitem{Luke:1992cs}
  M.E.~Luke and A.V.~Manohar,
  Phys. Lett. B {\bf 286} (1992) 348.

\bibitem{Fleming:1999ee}
  S.~Fleming, T.~Mehen, and I.W.~Stewart,
  Nucl. Phys. A {\bf 677} (2000) 313.

\bibitem{Stetcu:2010xq}
  I.~Stetcu, J.~Rotureau, B.R.~Barrett, and U.~van Kolck,
  Annals Phys. {\bf 325} (2010) 1644.

\bibitem{Fewster:1994sd}
  C.J.~Fewster,
  J. Phys. A {\bf 28} (1995) 1107.

\bibitem{Phillips:1996ae}
  D.R.~Phillips and T.D.~Cohen,
  Phys. Lett. B {\bf 390} (1997) 7.

\bibitem{Peierls:1959}
R.E. Peierls,
Proc. Roy. Soc. (London) A {\bf 253} (1959) 16. 

\bibitem{Hu:1948zz}
  N. Hu,
  Phys. Rev. {\bf 74} (1948) 131.

\bibitem{Schuetzer:1951}
W. Sch\"utzer and J. Tiomno,
Phys. Rev. {\bf 83} (1951) 249.

\bibitem{DemkovDrukarev1966}
Y.N. Demkov and G.F. Drukarev,
Sov. Phys. JETP {\bf 22} (1966) 479. 

\bibitem{Ma:1946}
  S.T. Ma,
  Phys. Rev. {\bf 69} (1946) 668.

\bibitem{TerHaar:1946}
  D. ter Haar,
  Physica {\bf 12} (1946) 501.

\bibitem{Ma:1947zz}
  S.T. Ma,
  Phys. Rev. {\bf 71} (1947) 195.

\bibitem{Nelson:1971nr}
C.~Nelson, A.~Rajagopal, and C.~Shastry,
J. Math. Phys. \textbf{12} (1971) 737.

\bibitem{Terry:1980bg}
P.~Terry,
J. Math. Phys. \textbf{23} (1982) 87.

\bibitem{Bargmann:1949zz}
  V. Bargmann,
  Phys. Rev. {\bf 75} (1949) 301.

\bibitem{Yamamoto:1962}
K. Yamamoto, 
Prog. Theor. Phys. {\bf 27} (1962) 219.

\bibitem{Mehen:1998zz}
  T. Mehen and I.W. Stewart,
  Phys. Lett. B {\bf 445} (1999) 378.

\bibitem{vanKampen:1953}
N.G.~van Kampen,
Phys. Rev. \textbf{91} (1953) 1267.

\bibitem{Nussenzveig:1959}
H.M. Nussenzveig,
Nucl. Phys. {\bf 11} (1959) 499.

\bibitem{Bedaque:1997qi}
P.F.~Bedaque and U.~van Kolck,
Phys. Lett. B \textbf{428} (1998) 221.

\bibitem{Bedaque:1998mb}
P.F.~Bedaque, H.-W. Hammer, and U.~van Kolck,
Phys. Rev. C \textbf{58} (1998) 641.

\bibitem{Bedaque:1998kg}
P.F.~Bedaque, H.-W.~Hammer, and U.~van Kolck,
Phys. Rev. Lett. \textbf{82} (1999) 463.

\bibitem{Bedaque:1998km}
P.F.~Bedaque, H.-W. Hammer, and U.~van Kolck,
Nucl. Phys. A \textbf{646} (1999) 444.

\bibitem{Bedaque:1999ve}
P.F.~Bedaque, H.-W.~Hammer, and U.~van Kolck,
Nucl. Phys. A \textbf{676} (2000) 357.

\end{thebibliography}
\end{document}